\documentclass[aps,pra,showpacs,twocolumn,nofootinbib]{revtex4-2}
\usepackage{hyperref}
\usepackage{bm}
\usepackage{amsmath}
\usepackage{amssymb}
\usepackage{slashed}
\usepackage{float}
\allowdisplaybreaks
\usepackage{longtable}
\usepackage{dcolumn}
\usepackage{graphicx}
\usepackage{xspace}
\usepackage{listings}
\usepackage{mathrsfs}

\usepackage{physics}
\usepackage{nicefrac}
\newcolumntype{w}[1]{D{.}{.}{#1}}
\newcommand{\Za}{Z\alpha}

\newcommand{\lbr}{\langle}
\newcommand{\rbr}{\rangle}

\newcommand{\KS}{K\"all{\'e}n-Sabry\xspace}

\newcommand{\vsp}{\mathbf}
\newcommand{\vspz}{\vsp{0}}
\newcommand{\vrel}{\mathit}
\newcommand{\schwinger}{\text{Schw}}
\newcommand{\ffdeg}{\mathrm{Deg}}

\begin{document}

\title{Two-loop vacuum polarization in a Coulomb field}

\author{S.~A. Volkov}
\affiliation{Max~Planck~Institute for Nuclear Physics, Saupfercheckweg~1, D~69117 Heidelberg, Germany}

\author{V.~A. Yerokhin}
\affiliation{Max~Planck~Institute for Nuclear Physics, Saupfercheckweg~1, D~69117 Heidelberg, Germany}

\author{Z. Harman}
\affiliation{Max~Planck~Institute for Nuclear Physics, Saupfercheckweg~1, D~69117 Heidelberg, Germany}

\author{C.~H. Keitel}
\affiliation{Max~Planck~Institute for Nuclear Physics, Saupfercheckweg~1, D~69117 Heidelberg, Germany}

\begin{abstract}

The leading-order two-loop vacuum-polarization potential, linear in the Coulomb field of a nucleus, was
first derived in the seminal 1955 work of \KS. The higher-order two-loop
vacuum-polarization corrections, however, have remained unknown until now.
In this work, we compute Coulomb corrections to the \KS potential, specifically those
involving three, five, and seven Coulomb interactions inside the vacuum-polarization loop.
The potentials are evaluated in momentum space and subsequently used to calculate one-electron energy shifts.
Our results reduce the theoretical uncertainty of the two-loop vacuum-polarization contribution
to transition energies, which is
required for next-generation tests of bound-state QED in heavy one and few-electron
ions as well as for the determination of nuclear charge radii.
\end{abstract}

\maketitle

\section{Introduction}

Theoretical and experimental investigations of the Lamb
shift in atomic systems provide stringent tests of bound-state quantum electrodynamics (QED)
theory, including the regime where the electron-nucleus coupling
parameter $\Za$ is not small
(where $Z$ is the nuclear charge number and $\alpha$ is the fine-structure constant)
\cite{indelicato:19}.
QED effects of first order in $\alpha$ have been extensively calculated and verified
by numerous experiments \cite{mohr:98}. In contrast,
calculations of two-loop QED effects, of order $\alpha^2$, are not
fully completed yet.
In hydrogen-like atoms, the two-loop effects typically determine the theoretical
uncertainty of energy levels and thus represent a central challenge in advancing our
understanding of these systems \cite{yerokhin:18:hydr}.
Similarly, in few-electron highly charged ions, two-loop QED corrections contribute significantly
to the theoretical uncertainties, thereby limiting comparisons with available experimental data
\cite{beiersdorfer:93,beiersdorfer:95} and possible determinations of nuclear charge radii
\cite{yerokhin:25:Lilike:arxiv}.

The two-loop QED effects have been extensively investigated during the last decades, both within
methods based on the $\Za$ expansion
\cite{czarnecki:05:prl,jentschura:05:sese,dowling:10,czarnecki:16} and also within the
all-order (in $\Za$) approach \cite{yerokhin:03:prl,yerokhin:06:prl}.
In particular, the two-loop self-energy correction has been evaluated nonperturbatively
in $\Za$
\cite{yerokhin:24:sese,yerokhin:25:sese} as well as most of two-loop diagrams with
vacuum-polarization loops \cite{yerokhin:08:twoloop}.
The largest uncertainty of the two-loop QED contribution for hydrogen-like atoms
comes \cite{yerokhin:18:hydr} from two uncalculated effects:
(i) Coulomb corrections to the two-loop vacuum polarization
and (ii) the electron self-energy with a light-by-light-scattering
insertion into the photon line.
The goal of the present investigation is to compute the first of the two missing
effects, which is an important step towards completing
the long-standing project of calculation of the full set
of one-electron two-loop QED diagrams.

The vacuum-polarization (VP) potential can be conveniently represented
\cite{blomqvist1972vacuum} in the form of
a double expansion in $\alpha$ and $\Za$,
\begin{align}\label{eq:1}
V_{\rm VP}(\vsp{r}) = \sum_{i,j} V_{ij}(\vsp{r}) \equiv \sum_{{i=1,2,\ldots}\atop
{j = 1,3,\ldots}}
\alpha^i\,(\Za)^j \, \widetilde{V}_{ij}(\vsp{r})\,.
\end{align}
According to the Furry theorem, the expansion terms with even powers of $\Za$ vanish, so the
summation over $j$ includes only odd $j$'s.

At the one-loop ($i = 1$) level, the VP potential has been extensively studied in the literature.
The first term of the expansion, $V_{11}(\vsp{r})$, is the well-known Uehling potential \cite{uehling:35}.
The next-order term, $V_{13}(\vsp{r})$,
was derived by Wichmann and Kroll \cite{wichmann:56},
with its explicit coordinate-space form obtained in Ref.~\cite{blomqvist1972vacuum}.
The Coulomb corrections of order $(\Za)^3$ and higher to the one-loop VP are commonly referred to as
the Wichmann-Kroll potential,
which has been calculated to all orders in $\Za$ numerically
\cite{soff:88:vp,manakov:89:zhetp,persson:93:vp}. For the point nuclear model,
accurate approximate formulas for the Wichmann-Kroll potential are available in the
literature
\cite{fainshtein:91,manakov:12:vgu}.

The two-loop ($i = 2$) VP potential has been studied to a significantly lesser extent.
Only the leading $\Za$-expansion term, $V_{21}$ --
commonly referred to as the \KS potential --
is presently known \cite{kaellen:55}.
Explicit formulas for this potential can be found in
Refs.~\cite{blomqvist1972vacuum,fullerton:76}.
Some partial results for higher-order VP energy shifts were reported in Ref.~\cite{plunien:98:epj}.
In the present investigation we compute the Coulomb corrections to the \KS potential
of order $(\Za)^3$, $(\Za)^5$, and $(\Za)^7$,
namely, potentials ${V}_{23}$ , ${V}_{25}$, and ${V}_{27}$.


It is important to note that the $\Za$-expansion of the VP potential in Eq.~(\ref{eq:1})
remains useful not only when the parameter $\Za$ is small, but even as it approaches unity, as is the case for heavy ions.
For example, in the case of the $1s$ state of uranium ($Z = 92$), $V_{11}$ induces an
energy shift of $-94$~eV, $V_{13}$ contributes $4.6$~eV,
$V_{15}$ contributes $0.57$~eV, with the net effect of the remaining one-loop VP tail amounting to merely $0.17$~eV.
This is in contrast to the $\Za$ expansion of {\em energy shifts}, which converge poorly in the high-$Z$ regime.
For example, for the  $1s$ state of uranium, the leading $\Za$-expansion term gives $-64$~eV, significantly deviating from the full result
of $-89$~eV.
With this in mind, we will adopt the $\Za$ expansion of the potential in Eq.~(\ref{eq:1}) 
for our calculations of the two-loop VP potential.
This approach is supposed to remain applicable for most atoms of practical interest, with the exception of superheavy elements.

An additional advantage of using the expansion of Eq.~(\ref{eq:1}) is that the potentials $ \widetilde{V}_{ij}$
do not depend on $Z$  and, indeed, on any other parameters.
As a result, the potentials $\widetilde{V}_{23}$, $\widetilde{V}_{25}$, and $\widetilde{V}_{27}$, once computed and stored on a grid, can be employed for
evaluating energy shifts in arbitrary atoms, including muonic, antiprotonic, and other exotic atoms.

The paper is organized as follows. In Sec.~\ref{sec:method} we describe the method of calculation
of the VP potentials in the momentum space.
Sec.~\ref{sec:energ} describes calculations of the corresponding energy shifts.
Sec.~\ref{sec:res} presents the numerical results obtained for the VP potential and the corresponding
energy shifts, and discusses the experimental consequences of the obtained results.

The relativistic units ($\hbar=c=m=1$) are used throughout this paper, where $m$ is the electron mass.
We use bold face ($\vsp{p}$) for three-vectors and italic style ($\vrel{p}$) for four-vectors.
Four-vectors have the form $\vrel{p}=(\vrel{p}_0,\vsp{p}).$
We also use the notation $\slashed{p}=p_{\mu}\gamma^{\mu}$; the tensor $g_{\mu\nu}$ corresponds to the signature $(+,-,-,-)$, and the Dirac matrices fulfill the condition $\gamma_{\mu}\gamma_{\nu}+\gamma_{\nu}\gamma_{\mu}=2g_{\mu\nu}$.
The Coulomb potential is $V_{\text{C}}(r)=-Z\alpha/r$ in coordinate space.

\section{CALCULATION OF POTENTIALS}
\label{sec:method}

\begin{figure}[h]
\includegraphics[width=70mm]{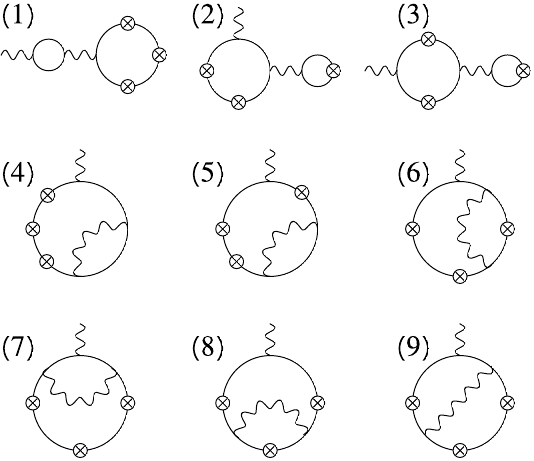}
\caption{Feynman diagrams contributing to $ {V}_{23}$.
Solid lines denote free-electron propagators,
wave lines denote the photon propagators, the circled crosses
denote the Coulomb interactions with the nucleus.
\label{fig23}
}
\end{figure}

\begin{figure}[h]
\includegraphics[width=85mm]{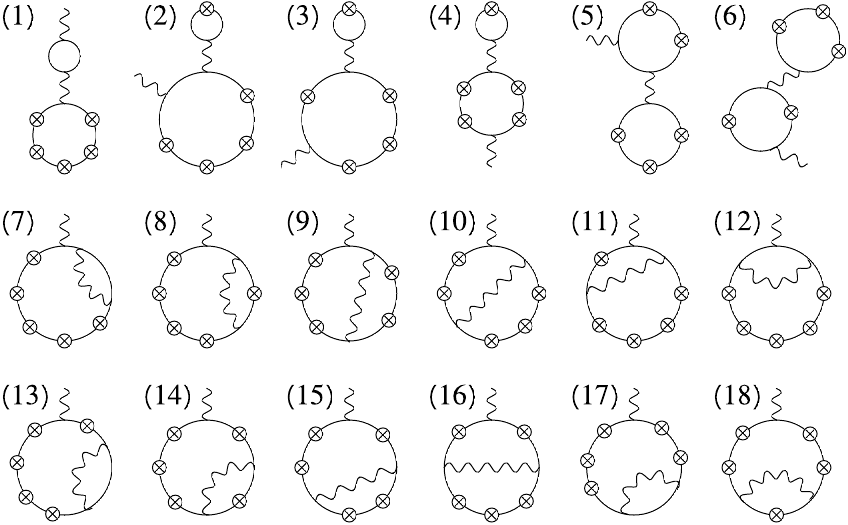}
\caption{Feynman diagrams contributing to $ {V}_{25}$.
\label{fig25}
}
\end{figure}

\begin{figure}[h]
\includegraphics[width=85mm]{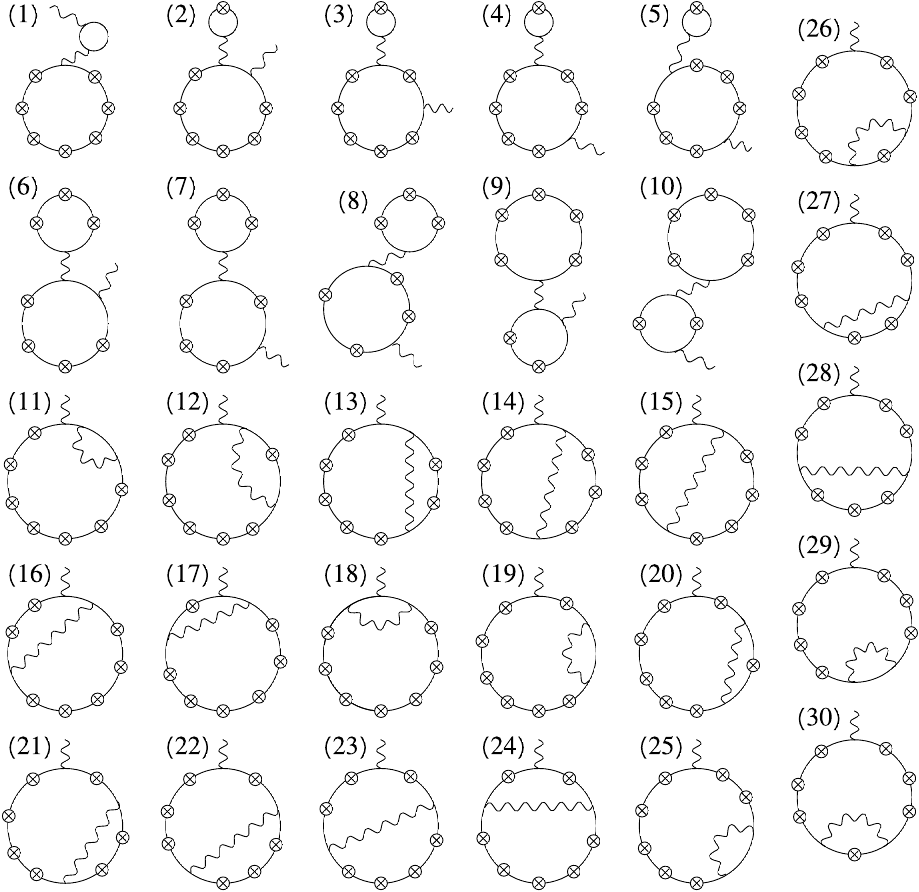}
\caption{Feynman diagrams contributing to $ {V}_{27}$.
\label{fig27}
}
\end{figure}

We calculate here the potentials ${V}_{23}$, ${V}_{25}$, and ${V}_{27}$.
The calculation is performed in momentum space, with the Fourier transform of the potential defined by
$$
 {V}(\vsp{p})=\int e^{-i\vsp{p}\cdot\vsp{r}}\,  {V}(\vsp{r})\, d^3 \vsp{r}.
$$
The Feynman diagrams contributing to ${V}_{23}$, ${V}_{25}$, and ${V}_{27}$
are shown in Fig.~\ref{fig23}, Fig.~\ref{fig25}, and Fig.~\ref{fig27}, respectively.
The set of Feynman diagrams for $ {V}_{ij}$ is obtained by taking all possible
graphs that satisfy the following conditions:
\begin{itemize}
\item they are connected;
\item they have one external photon line, $j$ Coulomb-potential insertions, no other external lines;
\item they have $i$ independent loops;
\item they do not have electron loops with odd number of vertices (Furry's theorem).
\end{itemize}
In our approach, the graphs (2)-(9)
in Fig.~\ref{fig23}, (2)-(18) in Fig.~\ref{fig25}, and (2)-(30) in Fig.~\ref{fig27} will be treated as genuine two-loop diagrams,
while the graphs (1) in all these figures are
expressed as products of two one-loop diagrams in momentum space
and evaluated separately.

\begin{figure}
\includegraphics[width=85mm]{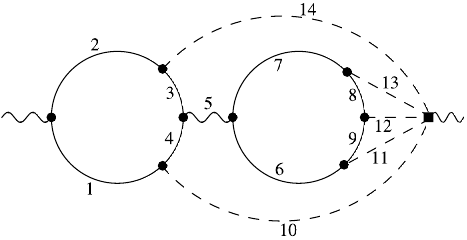}
\caption{The unfolded Feynman diagrams corresponding to the graph (6) in Fig. \ref{fig25}.}
\label{figdisclose}
\end{figure}

We employ methods of free QED in our calculation.
To this end, we first \emph{unfold} each Feynman diagram.
Unfolding means replacing each Coulomb-potential insertion with a Coulomb-propagator
line extending to an artificially-added vertex,
see Fig.~\ref{figdisclose} for an example.
In the figure, the additional vertex is shown as a black square dot,
and the dashed lines represent the Coulomb propagators.
The new vertex has a fictitious external line that carries away the excess momentum, so that momentum conservation holds
for all vertices.
The additional vertex contributes no multiplier and carries no tensor index.
In the Feynman gauge, the photon, electron, and Coulomb propagators are (up to  prefactors)
\begin{equation}\label{eq_propagators}
\frac{i(\slashed{\vrel{q}}+m)}{\vrel{q}^2-m^2+i0},\quad \frac{-ig_{\mu\nu}}{\vrel{q}^2+i0},\quad \frac{-\delta(\vrel{q}_0)\delta_{\mu 0}}{\vsp{q}^2},
\end{equation}
respectively.
The unfolded diagrams for $ {V}_{23}$, $ {V}_{25}$, and $ {V}_{27}$ have 4, 6, and 8 loops, respectively.
The large number of loops makes the computation of these Feynman diagrams demanding and time consuming.

In this work, 
we do not employ dimensional regularization or any other schemes based on expansions around singular points, which are commonly used in the literature.
Instead, we remove all divergences by performing point-by-point subtractions in the Feynman parametric space, 
by modifying integrands with suitably chosen counterterms that make the integrals finite. Point-by-point subtraction methods
are relatively rarely used in quantum field calculations, because of the complicated structure of divergences in Feynman diagrams.
However, they become advantageous in numerical computations of diagrams with many loops.
These methods were developed for calculations of
the free electron anomalous magnetic moment ~\cite{levinewright, carrollyao, kinoshita_6, kinoshita_10_first, volkov_5loops_total, kinoshita_verification_2025}; developments for other problems also exist \cite{kompaniets}.

Our method is based on a variant of the Bogoliubov-Parasiuk-Hepp-Zimmermann (BPHZ) renormalization method~\cite{bogolubovparasuk,hepp,zimmerman}. Applications of BPHZ to the two-loop QED effects in a Coulomb field have been previously studied in Ref.~\cite{zschocke_bphz_2002}. 
An advantage of our approach is that it directly reduces the problem  to numerically computable integrals. Our general approach
is based on a direct application of
Zimmermann's forest formula~\cite{zavialovstepanov,zimmerman}. This formula implements renormalization ``in place'' by adding counterterms to each Feynman amplitude. According to Zimmermann's forest formula, the renormalized Feynman amplitude is defined as follows:
\begin{itemize}
\item A summation over forests $F$ of ultraviolet (UV) divergent subdiagrams is performed; a forest is a set of subdiagrams, where each pair of subdiagrams is either nested or non-intersecting.
\item In each term, we replace the Feynman amplitude of each subdiagram $G'\in F$ with a momentum-space polynomial whose degree corresponds to the UV degree of divergence of $G'$, obtained by applying suitable projections. The transformation is organized so that it is first applied to smaller subgraphs (with respect to inclusion), and then carried over to larger subgraphs, ultimately yielding the final amplitude.  In the original formulations of the forest formula ~\cite{zavialovstepanov,zimmerman}, Taylor expansions around zero momenta are used; here, however, we employ slightly different projectors for certain types of subgraphs.
\item The coefficient of the term is $(-1)^{|F|}$.
\end{itemize}
 With $N_l$, $N_{\gamma}$ denoting the numbers of external electron and photon (or Coulomb) lines of a subdiagram, the UV-divergent subdiagrams in QED are: \emph{electron self-energy} ($N_l=2$, $N_{\gamma}=0$), \emph{vertex-like} ($N_l=2$, $N_{\gamma}=1$), \emph{photon self-energy} ($N_l=0$, $N_{\gamma}=2$), and \emph{photon-photon scattering} ($N_l=0$, $N_{\gamma}=4$). All such subdiagrams of the original Feynman diagram must be included to the forest formula. However, additional subdiagrams must also be included in the forest formula for the unfolded diagram, namely those with exactly two photon external lines (one fictitious) and no other external lines (except those connected to a fictitious vertex). These subdiagrams correspond to lower-order potential insertions and are also UV divergent. For example, in Fig. \ref{figdisclose} the following subdiagrams must be included: $\{1,2,3,4\}$, $\{6,7,8,9\}$, $\{6,7,8,9,11,12,13\}$, $\{1,2,\ldots,14\}$, where subgraphs are encoded by sets of internal graph lines. The full diagram is always included. In Zimmermann's forest formula, we take the value at zero momenta for these special subdiagrams, as well as for photon-photon scattering subgraphs. All subtraction contributions vanish identically due to gauge invariance once all Feynman diagrams are summed.

The subtraction projectors for the electron self-energy and vertex-like subdiagrams in Zimmermann's forest formula are usually
chosen to correspond to QED on-mass-shell renormalization conditions. The drawback of this choice is that
these conditions generate infrared (IR) divergences\footnote{See, e.g., the explicit one-loop expression for the renormalization constant (119.7a) in ~\cite{ll4}, Chapter XII (Radiative Corrections), Section 119 (Calculation of the Mass Operator).}.
We avoid IR divergences by using the projector $U$ defined as
$$
(U\Gamma)_{\mu}(\vrel{p},\vrel{q})=a((m_0)^2) \gamma_{\mu},
$$
$$
\quad (U\Sigma)(\vrel{p})=u(m^2)+v(m^2)m + v((m_0)^2)(\slashed{\vrel{p}}-m),
$$
where
$\Gamma_{\mu}(\vrel{p},\vrel{q})$ is a vertex-like Feynman amplitude,
$\vrel{p}-\frac{\vrel{q}}{2}$ and $\vrel{p}+\frac{\vrel{q}}{2}$ are the incoming and outgoing electron momenta, respectively,
$\vrel{q}$ is the photon momentum,
$$
\Gamma_{\mu}(\vrel{p},0)=a(\vrel{p}^2)\gamma_{\mu} + b(\vrel{p}^2)\vrel{p}_{\mu} + c(\vrel{p}^2)\slashed{\vrel{p}}\vrel{p}_{\mu} + d(\vrel{p}^2)(\slashed{\vrel{p}}\gamma_{\mu}-\gamma_{\mu}\slashed{\vrel{p}}),
$$
$\Sigma(\vrel{p})$ is an electron self-energy Feynman amplitude,
$$
\Sigma(\vrel{p})=u(\vrel{p}^2)+v(\vrel{p}^2)\slashed{\vrel{p}},
$$
and $(m_0)^2$ is a free parameter. In our calculations, we set
$$
(m_0)^2=-5m^2
$$
to minimize numerical cancellations inside integrals (this choice is motivated by our numerical experience).
Similar projectors were introduced previously for the free electron $g$-factor calculations: first in Ref.~\cite{volkov_2015} with $(m_0)^2=m^2$, and later in the current form in Ref.~\cite{volkov_acat_2021}.
The projector $U$ preserves the Ward identity, extracts the mass part completely at the physical point $m^2$, and does not generate IR divergences. It introduces an electron wave-function renormalization factor relative to standard QED on-mass-shell renormalization, but this factor does not appear in our case, as our Feynman diagrams have no external electron lines\footnote{In our case it can be checked immediately by collecting the corresponding differences in all Feynman diagrams, taking into account the one-loop Ward identity. The general-case statement is an important component of the QED renormalizability proof. See, for example, ~\cite{collins}, its Chapter 5 ``Renormalization'', Section 5.6 ``Relation to $\mathscr{L}$''. A clear explanation is also given in ~\cite{bogoliubov_shirkov}. See also an application example of changing the subtraction point in the Section 19.9 ``Analytic Continuation and Intermediate Renormalization'' of ~\cite{bjorkendrell2}.}.

\begin{table*}[htb]
\begin{center}
\caption{The values of $\widetilde{V}_{23}(\vsp{p})$ for different $|\vsp{p}|$
\label{table_v_23}
}
\begin{ruledtabular}
\begin{tabular}{cccccccc}
 $|\vsp{p}|$ & $\widetilde{V}_{23}(\vsp{p})$ & $|\vsp{p}|$ & $\widetilde{V}_{23}(\vsp{p})$ & $|\vsp{p}|$ & $\widetilde{V}_{23}(\vsp{p})$ & $|\vsp{p}|$ & $\widetilde{V}_{23}(\vsp{p})$ \\ \hline \\
0.001 & $0.062303(73)$ & 0.003 & $0.062197(69)$ & 0.005 & $0.062214(68)$ & 0.01 & $0.062164(67)$ \\
0.03 & $0.062011(66)$ & 0.05 & $0.062171(66)$ & 0.1 & $0.061785(64)$ & 0.15 & $0.061358(63)$ \\
0.16 & $0.061254(63)$ & 0.17 & $0.061148(63)$ & 0.18 & $0.061115(63)$ & 0.19 & $0.060879(62)$ \\
0.2 & $0.060828(62)$ & 0.22 & $0.060551(61)$ & 0.24 & $0.060286(61)$ & 0.26 & $0.059995(61)$ \\
0.28 & $0.059852(61)$ & 0.3 & $0.059321(60)$ & 0.32 & $0.059028(59)$ & 0.34 & $0.058693(59)$ \\
0.36 & $0.058397(58)$ & 0.38 & $0.057974(58)$ & 0.4 & $0.057655(57)$ & 0.42 & $0.057203(57)$ \\
0.44 & $0.056844(56)$ & 0.46 & $0.056455(54)$ & 0.48 & $0.056028(54)$ & 0.5 & $0.055616(53)$ \\
0.55 & $0.054473(52)$ & 0.6 & $0.053423(52)$ & 0.65 & $0.052335(50)$ & 0.7 & $0.051293(48)$ \\
0.75 & $0.049992(46)$ & 0.8 & $0.048804(45)$ & 0.85 & $0.047781(44)$ & 0.9 & $0.046518(43)$ \\
0.95 & $0.045357(42)$ & 1 & $0.044156(40)$ & 1.1 & $0.041927(37)$ & 1.2 & $0.039814(34)$ \\
1.3 & $0.037642(32)$ & 1.4 & $0.035669(30)$ & 1.5 & $0.033819(28)$ & 1.6 & $0.032027(27)$ \\
1.7 & $0.030337(27)$ & 1.8 & $0.028775(25)$ & 1.9 & $0.027324(22)$ & 2 & $0.025955(22)$ \\
2.1 & $0.024688(21)$ & 2.2 & $0.023483(20)$ & 2.3 & $0.022408(18)$ & 2.4 & $0.021361(17)$ \\
2.5 & $0.020364(16)$ & 2.6 & $0.019493(15)$ & 2.7 & $0.018654(15)$ & 2.8 & $0.017837(14)$ \\
2.9 & $0.017099(14)$ & 3 & $0.016403(13)$ & 3.5 & $0.013579(11)$ & 4 & $0.0114472(89)$ \\
4.5 & $0.0098321(77)$ & 5 & $0.0085598(65)$ & 5.5 & $0.0075518(60)$ & 6 & $0.0067090(51)$ \\
6.5 & $0.0060201(47)$ & 7 & $0.0054442(42)$ & 7.5 & $0.0049455(38)$ & 8 & $0.0045219(34)$ \\
9 & $0.0038329(29)$ & 10 & $0.0032963(25)$ & 11 & $0.0028723(22)$ & 12 & $0.0025336(19)$ \\
13 & $0.0022512(17)$ & 14 & $0.0020171(15)$ & 15 & $0.0018173(14)$ & 16 & $0.0016467(13)$ \\
18 & $0.0013783(10)$ & 20 & $1.17228(88)\times 10^{-3}$ & 22 & $1.00985(76)\times 10^{-3}$ & 24 & $8.8080(66)\times 10^{-4}$ \\
26 & $7.7702(59)\times 10^{-4}$ & 28 & $6.9022(52)\times 10^{-4}$ & 30 & $6.1699(47)\times 10^{-4}$ & 32 & $5.5609(42)\times 10^{-4}$ \\
36 & $4.5895(35)\times 10^{-4}$ & 40 & $3.8649(30)\times 10^{-4}$ & 44 & $3.3052(25)\times 10^{-4}$ & 48 & $2.8593(22)\times 10^{-4}$ \\
52 & $2.5017(19)\times 10^{-4}$ & 56 & $2.2067(17)\times 10^{-4}$ & 60 & $1.9661(15)\times 10^{-4}$ & 64 & $1.7654(14)\times 10^{-4}$ \\
72 & $1.4452(11)\times 10^{-4}$ & 80 & $1.20605(93)\times 10^{-4}$ & 88 & $1.02467(79)\times 10^{-4}$ & 96 & $8.8140(68)\times 10^{-5}$ \\
104 & $7.6793(59)\times 10^{-5}$ & 112 & $6.7465(52)\times 10^{-5}$ & 120 & $5.9828(46)\times 10^{-5}$ & 128 & $5.3474(41)\times 10^{-5}$ \\
144 & $4.3557(34)\times 10^{-5}$ & 160 & $3.6165(28)\times 10^{-5}$ & 176 & $3.0591(24)\times 10^{-5}$ & 192 & $2.6203(20)\times 10^{-5}$ \\
208 & $2.2756(18)\times 10^{-5}$ & 224 & $1.9979(16)\times 10^{-5}$ & 240 & $1.7617(14)\times 10^{-5}$ & 256 & $1.5720(12)\times 10^{-5}$ \\
288 & $1.2768(10)\times 10^{-5}$ & 320 & $1.05402(82)\times 10^{-5}$ & 352 & $8.8978(70)\times 10^{-6}$ & 384 & $7.6032(60)\times 10^{-6}$ \\
416 & $6.5813(52)\times 10^{-6}$ & 448 & $5.7475(45)\times 10^{-6}$ & 480 & $5.0872(40)\times 10^{-6}$ & 512 & $4.5276(35)\times 10^{-6}$ \\
1000 & $1.3361(10)\times 10^{-6}$ & 2000 & $3.7243(29)\times 10^{-7}$ & 5000 & $6.7753(54)\times 10^{-8}$ & 10000 & $1.8487(15)\times 10^{-8}$ \\
100000 & $2.3623(19)\times 10^{-10}$ & \ & \ & \ & \ & \ & \ \\
\end{tabular}
\end{ruledtabular}
\end{center}
\end{table*}

\begin{table*}[htb]
\begin{center}
\caption{The values of $\widetilde{V}_{25}(\vsp{p})$ for different $|\vsp{p}|$
\label{table_v_25}
}
\begin{ruledtabular}
\begin{tabular}{cccccccc}
 $|\vsp{p}|$ & $\widetilde{V}_{25}(\vsp{p})$ & $|\vsp{p}|$ & $\widetilde{V}_{25}(\vsp{p})$ & $|\vsp{p}|$ & $\widetilde{V}_{25}(\vsp{p})$ & $|\vsp{p}|$ & $\widetilde{V}_{25}(\vsp{p})$ \\ \hline \\
0.001 & $0.014832(67)$ & 0.003 & $0.014779(66)$ & 0.005 & $0.014825(66)$ & 0.01 & $0.014859(67)$ \\
0.03 & $0.015011(67)$ & 0.05 & $0.014838(66)$ & 0.1 & $0.014815(66)$ & 0.15 & $0.014764(66)$ \\
0.16 & $0.014836(66)$ & 0.17 & $0.014713(65)$ & 0.18 & $0.014739(65)$ & 0.19 & $0.014735(65)$ \\
0.2 & $0.014766(64)$ & 0.22 & $0.014605(64)$ & 0.24 & $0.014693(64)$ & 0.26 & $0.014628(64)$ \\
0.28 & $0.014656(64)$ & 0.3 & $0.014582(64)$ & 0.32 & $0.014589(64)$ & 0.34 & $0.014557(64)$ \\
0.36 & $0.014388(63)$ & 0.38 & $0.014435(63)$ & 0.4 & $0.014515(63)$ & 0.42 & $0.014308(62)$ \\
0.44 & $0.014204(61)$ & 0.46 & $0.014232(62)$ & 0.48 & $0.014094(61)$ & 0.5 & $0.014074(61)$ \\
0.55 & $0.013827(60)$ & 0.6 & $0.013734(60)$ & 0.65 & $0.013670(59)$ & 0.7 & $0.013371(57)$ \\
0.75 & $0.013176(56)$ & 0.8 & $0.013021(56)$ & 0.85 & $0.012758(54)$ & 0.9 & $0.012657(54)$ \\
0.95 & $0.012420(53)$ & 1 & $0.012235(52)$ & 1.1 & $0.011815(50)$ & 1.2 & $0.011402(48)$ \\
1.3 & $0.010959(46)$ & 1.4 & $0.010643(45)$ & 1.5 & $0.010117(42)$ & 1.6 & $0.009692(41)$ \\
1.7 & $0.009350(39)$ & 1.8 & $0.009028(37)$ & 1.9 & $0.008634(36)$ & 2 & $0.008362(34)$ \\
2.1 & $0.007971(33)$ & 2.2 & $0.007741(32)$ & 2.3 & $0.007468(30)$ & 2.4 & $0.007182(29)$ \\
2.5 & $0.006917(28)$ & 2.6 & $0.006681(27)$ & 2.7 & $0.006475(26)$ & 2.8 & $0.006293(25)$ \\
2.9 & $0.006058(24)$ & 3 & $0.005907(23)$ & 3.5 & $0.005059(20)$ & 4 & $0.004406(17)$ \\
4.5 & $0.003849(15)$ & 5 & $0.003451(13)$ & 5.5 & $0.003088(12)$ & 6 & $0.002776(10)$ \\
6.5 & $0.0025248(94)$ & 7 & $0.0023178(87)$ & 7.5 & $0.0021048(78)$ & 8 & $0.0019559(73)$ \\
9 & $0.0016760(61)$ & 10 & $0.0014618(53)$ & 11 & $0.0012837(47)$ & 12 & $0.0011450(42)$ \\
13 & $0.0010231(38)$ & 14 & $0.0009142(33)$ & 15 & $0.0008299(30)$ & 16 & $0.0007583(27)$ \\
18 & $0.0006360(23)$ & 20 & $0.0005433(20)$ & 22 & $0.0004716(17)$ & 24 & $0.0004152(15)$ \\
26 & $0.0003655(13)$ & 28 & $0.0003237(12)$ & 30 & $0.0002919(11)$ & 32 & $2.6274(95)\times 10^{-4}$ \\
36 & $2.1976(80)\times 10^{-4}$ & 40 & $1.8432(67)\times 10^{-4}$ & 44 & $1.5679(57)\times 10^{-4}$ & 48 & $1.3651(50)\times 10^{-4}$ \\
52 & $1.2011(44)\times 10^{-4}$ & 56 & $1.0711(39)\times 10^{-4}$ & 60 & $9.468(34)\times 10^{-5}$ & 64 & $8.541(31)\times 10^{-5}$ \\
72 & $6.975(25)\times 10^{-5}$ & 80 & $5.821(21)\times 10^{-5}$ & 88 & $4.988(18)\times 10^{-5}$ & 96 & $4.278(16)\times 10^{-5}$ \\
104 & $3.764(14)\times 10^{-5}$ & 112 & $3.290(12)\times 10^{-5}$ & 120 & $2.909(11)\times 10^{-5}$ & 128 & $2.612(10)\times 10^{-5}$ \\
144 & $2.1286(78)\times 10^{-5}$ & 160 & $1.7672(65)\times 10^{-5}$ & 176 & $1.5059(55)\times 10^{-5}$ & 192 & $1.2793(47)\times 10^{-5}$ \\
208 & $1.1140(41)\times 10^{-5}$ & 224 & $9.834(36)\times 10^{-6}$ & 240 & $8.645(31)\times 10^{-6}$ & 256 & $7.675(28)\times 10^{-6}$ \\
288 & $6.258(23)\times 10^{-6}$ & 320 & $5.161(19)\times 10^{-6}$ & 352 & $4.373(16)\times 10^{-6}$ & 384 & $3.759(14)\times 10^{-6}$ \\
416 & $3.233(12)\times 10^{-6}$ & 448 & $2.852(11)\times 10^{-6}$ & 480 & $2.5110(93)\times 10^{-6}$ & 512 & $2.2388(83)\times 10^{-6}$ \\
1000 & $6.649(25)\times 10^{-7}$ & 2000 & $1.8530(69)\times 10^{-7}$ & 5000 & $3.389(13)\times 10^{-8}$ & 10000 & $9.278(35)\times 10^{-9}$ \\
100000 & $1.1819(44)\times 10^{-10}$ & \ & \ & \ & \ & \ & \ \\
\end{tabular}
\end{ruledtabular}
\end{center}
\end{table*}

\begin{table*}[htb]
\begin{center}
\caption{The values of $\widetilde{V}_{27}(\vsp{p})$ for different $|\vsp{p}|$
\label{table_v_27}
}
\begin{ruledtabular}
\begin{tabular}{cccccccc}
 $|\vsp{p}|$ & $\widetilde{V}_{27}(\vsp{p})$ & $|\vsp{p}|$ & $\widetilde{V}_{27}(\vsp{p})$ & $|\vsp{p}|$ & $\widetilde{V}_{27}(\vsp{p})$ & $|\vsp{p}|$ & $\widetilde{V}_{27}(\vsp{p})$ \\ \hline \\
0.001 & $0.00824(14)$ & 0.003 & $0.00850(14)$ & 0.005 & $0.00814(15)$ & 0.01 & $0.00844(14)$ \\
0.03 & $0.00841(14)$ & 0.05 & $0.00866(15)$ & 0.1 & $0.00829(14)$ & 0.15 & $0.00814(14)$ \\
0.16 & $0.00825(14)$ & 0.17 & $0.00838(14)$ & 0.18 & $0.00850(14)$ & 0.19 & $0.00829(14)$ \\
0.2 & $0.00829(14)$ & 0.22 & $0.00850(14)$ & 0.24 & $0.00815(14)$ & 0.26 & $0.00829(14)$ \\
0.28 & $0.00809(14)$ & 0.3 & $0.00834(14)$ & 0.32 & $0.00820(13)$ & 0.34 & $0.00803(14)$ \\
0.36 & $0.00821(14)$ & 0.38 & $0.00809(13)$ & 0.4 & $0.00813(14)$ & 0.42 & $0.00804(14)$ \\
0.44 & $0.00801(14)$ & 0.46 & $0.00808(13)$ & 0.48 & $0.00800(13)$ & 0.5 & $0.00789(13)$ \\
0.55 & $0.00776(13)$ & 0.6 & $0.00764(13)$ & 0.65 & $0.00785(13)$ & 0.7 & $0.00786(13)$ \\
0.75 & $0.00753(13)$ & 0.8 & $0.00766(12)$ & 0.85 & $0.00746(12)$ & 0.9 & $0.00725(12)$ \\
0.95 & $0.00705(12)$ & 1 & $0.00708(12)$ & 1.1 & $0.00716(11)$ & 1.2 & $0.00672(11)$ \\
1.3 & $0.00654(11)$ & 1.4 & $0.00626(10)$ & 1.5 & $0.00608(10)$ & 1.6 & $0.005817(94)$ \\
1.7 & $0.005681(92)$ & 1.8 & $0.005497(89)$ & 1.9 & $0.005420(85)$ & 2 & $0.005154(82)$ \\
2.1 & $0.004938(78)$ & 2.2 & $0.004875(76)$ & 2.3 & $0.004726(75)$ & 2.4 & $0.004564(71)$ \\
2.5 & $0.004348(68)$ & 2.6 & $0.004215(68)$ & 2.7 & $0.004190(68)$ & 2.8 & $0.003984(64)$ \\
2.9 & $0.003957(63)$ & 3 & $0.003736(59)$ & 3.5 & $0.003347(52)$ & 4 & $0.002932(45)$ \\
4.5 & $0.002692(42)$ & 5 & $0.002322(35)$ & 5.5 & $0.002033(31)$ & 6 & $0.001873(28)$ \\
6.5 & $0.001700(24)$ & 7 & $0.001556(23)$ & 7.5 & $0.001466(22)$ & 8 & $0.001328(19)$ \\
9 & $0.001165(17)$ & 10 & $0.001011(15)$ & 11 & $0.000877(13)$ & 12 & $0.000791(12)$ \\
13 & $0.000722(10)$ & 14 & $0.0006303(91)$ & 15 & $0.0005762(85)$ & 16 & $0.0005149(73)$ \\
18 & $0.0004433(64)$ & 20 & $0.0003773(52)$ & 22 & $0.0003286(49)$ & 24 & $0.0002928(41)$ \\
26 & $0.0002540(38)$ & 28 & $0.0002278(33)$ & 30 & $0.0002082(29)$ & 32 & $0.0001805(27)$ \\
36 & $0.0001510(21)$ & 40 & $0.0001281(19)$ & 44 & $0.0001079(16)$ & 48 & $0.0000945(14)$ \\
52 & $0.0000833(12)$ & 56 & $0.0000729(11)$ & 60 & $6.405(91)\times 10^{-5}$ & 64 & $6.026(86)\times 10^{-5}$ \\
72 & $4.758(71)\times 10^{-5}$ & 80 & $4.023(59)\times 10^{-5}$ & 88 & $3.485(49)\times 10^{-5}$ & 96 & $2.987(43)\times 10^{-5}$ \\
104 & $2.533(36)\times 10^{-5}$ & 112 & $2.300(33)\times 10^{-5}$ & 120 & $2.042(29)\times 10^{-5}$ & 128 & $1.820(26)\times 10^{-5}$ \\
144 & $1.468(21)\times 10^{-5}$ & 160 & $1.212(18)\times 10^{-5}$ & 176 & $1.032(15)\times 10^{-5}$ & 192 & $8.90(13)\times 10^{-6}$ \\
208 & $7.90(12)\times 10^{-6}$ & 224 & $6.91(10)\times 10^{-6}$ & 240 & $5.948(91)\times 10^{-6}$ & 256 & $5.390(78)\times 10^{-6}$ \\
288 & $4.351(65)\times 10^{-6}$ & 320 & $3.575(53)\times 10^{-6}$ & 352 & $2.955(43)\times 10^{-6}$ & 384 & $2.592(38)\times 10^{-6}$ \\
416 & $2.305(35)\times 10^{-6}$ & 448 & $1.990(29)\times 10^{-6}$ & 480 & $1.763(26)\times 10^{-6}$ & 512 & $1.565(23)\times 10^{-6}$ \\
1000 & $4.596(70)\times 10^{-7}$ & 2000 & $1.261(19)\times 10^{-7}$ & 5000 & $2.348(36)\times 10^{-8}$ & 10000 & $6.271(94)\times 10^{-9}$ \\
100000 & $8.10(12)\times 10^{-11}$ & \ & \ & \ & \ & \ & \ \\
\end{tabular}
\end{ruledtabular}
\end{center}
\end{table*}

Our method yields a single Feynman-parametric integral
\begin{equation}\label{eq_feyn_param_integral}
\int_{z\geq 0} F(\vsp{p},z_1,\ldots,z_n)\delta(z_1+\ldots+z_n-1) d^n z
\end{equation}
for each Feynman diagram.
Here, $z_j$ corresponds to the $j$th internal line of the \emph{unfolded} diagram,
 and $\vsp{p}$ is the external momentum. These integrals are finite and can be computed separately for each diagram and then summed.
Our procedure for obtaining the functions $F$ involves the Schwinger parametrization as an intermediate step, as follows:
\begin{itemize}
\item
Introduce the Schwinger parametrization for the propagators in Eq.~(\ref{eq_propagators}), obtaining
\begin{eqnarray*}
& (\slashed{\vrel{q}}+m)e^{iz_j(\vrel{q}^2-m^2+i\varepsilon)},\quad ig_{\mu\nu}e^{iz_j (\vrel{q}^2+i\varepsilon)},&
\\
&i\delta(\vrel{q}_0)\delta_{\mu 0} e^{iz_j(\vsp{q}^2+i\varepsilon)},&
\end{eqnarray*}
where $z_1,\ldots,z_n\geq 0$ are the Schwinger parameters, $\varepsilon>0$ is an infrared regulator, and $j$ labels the diagram line.

\item
Perform the loop-momentum integrations. We assume the forest formula described above is applied, but \emph{the whole diagram is excluded from it}.
The momentum integral diverges, but at this step we ignore the divergence and use the standard formulas for multidimensional Gaussian integrals with polynomial factors. We obtain
\begin{eqnarray*}
&F^{\schwinger}(\vsp{p},z_1,\ldots,z_n,\varepsilon)=\frac{1}{D_0(z)^{1/2}D_1(z)^{3/2}}&
\\
&\times \left[ \sum_{l=0}^{s/2} R_l(\vsp{p}^2,z) \right] e^{iA(z)\vsp{p}^2+iB(z)-\varepsilon \sum_{j=1}^n z_j}&
\\
&+\text{counterterms},&
\end{eqnarray*}
where $D_0$ and $D_1$ are the first Symanzik polynomials\footnote{See, for example, Ref. ~\cite{bogner_weinzierl_feyn_graph_poly}.} of the original and unfolded diagrams, respectively; $R_l$ is a polynomial in $\vsp{p}^2$ (constant for $l=s/2$), rational and homogeneous of degree $-l$ in $z$; the symbol $s$ denotes the number of electron lines in the diagram; $A$ and $B$ are rational, homogeneous functions of degree $1$, and real; counterterms have the same form, but with different $D_j,R_j,A$; see, for example, ~\cite{bogoliubov_shirkov,zavialov,smirnov}.

\item Switch to Feynman parameters by
\begin{align}
F(\vsp{p},z)=& \frac{ 1}{\vsp{p}^2}\lim_{\varepsilon\rightarrow +0} \int_0^{+\infty} \lambda^{n-1}
\nonumber
\\
 \times & \left[ F^{\schwinger}(\vsp{p},\lambda z,\varepsilon)-F^{\schwinger}(\vspz,\lambda z,\varepsilon)\right] d\lambda\,.
\end{align}
where the denominator $\vsp{p}^2$ originates from the propagator of the external photon (which must be included in the potential as an internal photon propagator), while the subtraction of the $\vsp{p}=\vspz$ term implements the remaining subtraction of the whole diagram in Zimmermann's forest formula.
\end{itemize}

The integration over $\lambda$ and the limit can be performed analytically, yielding
\begin{align}
F(\vsp{p},z) =& \frac{1}{D_0(z)^{1/2}D_1(z)^{3/2}}
\nonumber \\
&\times\left\{ \frac{W_0(z)}{\vsp{p}^2} \log\left(1+\frac{A(z)}{B(z)}\vsp{p}^2\right) \right.
\nonumber \\
&+\sum_{l=1}^{s/2} \left[ \frac{W_l(z)}{\vsp{p}^2} \left( \frac{1}{(A(z)\vsp{p}^2+B(z))^l}-\frac{1}{B(z)^l} \right)\right]
\nonumber \\
&\left. +\sum_{l=1}^{s/2} \frac{Y_l(\vsp{p}^2,z)}{(A(z)\vsp{p}^2+B(z))^l}\right\}+\text{counterterms}\,,
\nonumber
\end{align}
where $W_l(z)$ are rational in $z$ and $Y_l(\vsp{p}^2,z)$ are polynomial in $\vsp{p}^2$ and rational in $z$. Each $W_l(z)$ is easily extractable from the constant term of $R_{\frac{s}{2}-l}(\vsp{p}^2,z)$ in $\vsp{p}^2$,
while each $Y_l(\vsp{p}^2,z)$ comes from the remaining terms. The $\log$-term arises from the logarithmic overall UV divergence of the full diagram; the separate subtraction ensures that these $\log$-terms are finite.

The value $F(\vspz,z)$ can be easily obtained by taking the limit $\vsp{p}\rightarrow \vspz$. However, the corresponding integral (\ref{eq_feyn_param_integral}) may diverge at $\vsp{p}=\vspz$. The total value of ${V}_{ij}(\vsp{p}=\vspz)$ is finite, but in our calculation we can't guarantee explicitly finite individual integrals at $\vsp{p}=\vspz$.
As a consequence, we are able to perform computations for small nonzero values of $|\vsp{p}|$, but not at $\vsp{p}=\vspz$.
However, extrapolation to $\vsp{p}=\vspz$ suffices for all practical purposes.


We evaluate the integrals (\ref{eq_feyn_param_integral}) numerically using the Monte Carlo method. It is well known that the convergence rate of Monte Carlo integration depends critically on the choice of the probability density function (PDF). Adaptive algorithms that fit the PDF to the integrand exist in the literature, but they can only adjust a relatively small number of parameters. For the complicated structure of typical integrands in Feynman-parameter integrations, such adaptive algorithms become inefficient at high dimensionality. In the present work, we employ a nonadaptive algorithm based on a theoretical analysis of the integrand's behavior  in Feynman-parametric space. This algorithm provides a unique PDF for each Feynman diagram. The problem of constructing this PDF is closely related to constructing
a good upper bound on the integrand absolute value and to proving the finiteness of renormalized Feynman amplitudes.
We borrow several ideas from these analyses.

To construct the PDF, we partition the integration domain into so-called Hepp sectors \cite{hepp}:
$$
z_{j_1}\geq z_{j_2}\geq \ldots \geq z_{j_n}\,.
$$
Each ordering of the Feynman parameters
defines one Hepp sector.
For example, for ${V}_{27}$ we have $n=18$, yielding $18!=6402373705728000$ sectors per diagram.

Our PDFs take the form
$$
C\times \frac{\prod_{l=2}^n \left( z_{j_l}/z_{j_{l-1}} \right)^{\ffdeg(\{j_l,j_{l+1},\ldots,j_M\})}}{z_1 \times z_2 \times \ldots \times z_n} + \text{stability terms},
$$
where $\ffdeg(s)$ are positive numbers defined on
all nonempty proper subsets of the set of all internal lines of the diagram.
For example, for ${V}_{27}$ we have $2^{18}-2=262142$ such numbers per diagram.
Functions of similar form were first used in Ref.~\cite{speer}. The necessity of  stability terms and their general construction were discussed in Ref.~\cite{volkov_gpu}, along with other stabilization techniques.
Efficient random-sample generation with such PDFs was described in Ref.~\cite{volkov_prd}; a similar algorithm was also implemented in \texttt{feyntrop} ~\cite{borinsky_quadrature,borinsky_minkowsky}.
The procedure for obtaining $\ffdeg(s)$ and stabilization terms, as well as further details, will be presented elsewhere.

The most resource-intensive part of the computation is the Monte Carlo integration itself. We employed NVidia P100 GPUs (accompanied by one Intel Xeon E5-2698 CPU core per GPU). Since all divergences are cancelled numerically under the integral sign, the subtractions may lead to large round-off errors. To control these, we used the interval arithmetic: 
at each arithmetic operation, we generate an interval $[x_1, x_2]$ that is guaranteed to contain the exact result $x$.
We employed five types of interval arithmetic that differ in speed and interval widths. 
We begin with a fast version of \texttt{double}-precision\footnote{Note that NVidia GPUs allow specifying rounding policies in arithmetic operations and also provide a precision-preserving realization of $\log(1+x)$.} interval arithmetic, which distributes arithmetic operations into blocks and estimates the interval width in each block without computing intervals for single operations \cite{volkov_gpu}. We then switch to the conventional interval arithmetic with  \texttt{double} precision.
If the achieved precision is insufficient, the same sample is recomputed with a higher-precision arithmetic.
Arbitrary-precision arithmetics with mantissas of 128, 256, and 384 bits were used. High-precision arithmetic is especially crucial for small values of $|\vsp{p}|$. The computations were performed on the MPIK Heidelberg computing cluster and required 15 GPU-days for each of $ {V}_{23}$, $ {V}_{25}$, and 42 GPU-days for $ {V}_{27}$. The corresponding numbers of Monte Carlo samples were $9\times 10^{12}$, $2.3\times 10^{12}$, and $6.1\times 10^{11}$, respectively. The compiled integrand code sizes were 30 MB for $ {V}_{23}$, 450 MB for $ {V}_{25}$, and 11 GB for $ {V}_{27}$.

\section{Calculation of energy shifts}
\label{sec:energ}

Once numerical values of the VP potentials are available, the corresponding
corrections to energy levels can be computed as expectation values of the potentials with the Dirac wave function of the
reference state. Since the potentials are obtained
in the momentum representation in this work, it is
natural to evaluate the expectation values also in momentum space,
\begin{align}
\langle V_{ij}\rangle
= \int \frac{d^3\vsp{p}_1}{(2\pi)^3}\,
\int \frac{d^3\vsp{p}_2}{(2\pi)^3} \,
\psi_a^{\dag}(\vsp{p}_1)\,
V_{ij}(\vsp{p}_1-\vsp{p}_2)\,
\psi_a(\vsp{p}_2)\,.
\end{align}
Integration over most angular variables is easily performed analytically
(see, e.g., Sec.~2C of Ref.~\cite{yerokhin:25:se}),
leaving a
three-dimensional integral to be carried out numerically
\begin{align}
\langle V_{ij}\rangle
= &\
\frac1{32\pi^5}
\int_0^{\infty} dp_1\,dp_2 \int_{-1}^1 d\xi\,
 (p_1p_2)^2\,
 V_{ij}(q)\,
 \nonumber \\  & \times
 \Big[ g_a(p_1)\,g_a(p_2)\,P_{l_a}(\xi) + f_a(p_1)\,f_a(p_2)\,P_{\overline{l}_a}(\xi) \Big]
 \,,
\end{align}
where $p_1 = |\vsp{p}_1|$, $p_2 = |\vsp{p}_2|$, $q = |\vsp{p}_1-\vsp{p}_2|$,
\mbox{$\xi = \vsp{p}_1\cdot\vsp{p}_2/(p_1p_2) $},
$P_l$ is the Legendre polynomial, $l_a = |\kappa_a+1/2|-1/2$, $\overline{l}_a = |\kappa_a-1/2|-1/2$,
and $g_a(p)$ and $f_a(p)$ are the upper and lower radial components of the Dirac wave
function, respectively, defined as in Ref.~\cite{yerokhin:25:se}.

For performing the numerical integrations, it is convenient to make the change variables
\cite{yerokhin:99:pra}
$(p_1,p_2,\xi) \to (x,y,q)$,
where $x = p_1+p_2$ and $y = |p_1-p_2|$, with the result
\begin{align}
\langle V_{ij}\rangle
= &\
\frac1{32\pi^5}
\int_0^{\infty} dx
\int_0^x dy\,
\int_y^x dq\,
 p_1p_2q\,
  V_{ij}(q)\,
 \nonumber \\  & \times
 \Big[ g_a(p_1)\,g_a(p_2)\,P_{l_a}(\xi) + f_a(p_1)\,f_a(p_2)\,P_{\overline{l}_a}(\xi) \Big]
 \,.
\end{align}

\section{Results}
\label{sec:res}

Our numerical results obtained for the potentials $\widetilde{V}_{23}(\vsp{p})$, $\widetilde{V}_{25}(\vsp{p})$, and $\widetilde{V}_{27}(\vsp{p})$ are presented in
Tables~\ref{table_v_23}, \ref{table_v_25}, and \ref{table_v_27}, respectively. The specified uncertainties
originate from Monte-Carlo integrations and represent $1\sigma$ deviations. The uncertainties for
different $\vsp{p}$'s are statistically independent. For small $\vsp{p}$ our results for
the potential $\widetilde{V}_{23}(\vsp{p})$ are in good agreement with the analytical result
by Krachkov and Lee \cite{lee_krachkov_2023},
$
\widetilde{V}_{23}(\vspz)=0.062\,214\ldots\,.
$
As a cross-check of our method and numerical procedure, we also calculated the one-loop potentials $\widetilde{V}_{13}(\vsp{p})$, $\widetilde{V}_{15}(\vsp{p})$, and $\widetilde{V}_{17}(\vsp{p})$,
and the two-loop potential $\widetilde{V}_{21}(\vsp{p})$,
with numerical values
presented in Supplementary Material. They were shown to be in excellent agreement with
a numerical Fourier transform of the corresponding coordinate-space potentials, see Ref.~\cite{yerokhin:08:twoloop} for details.

\begin{figure*}
\includegraphics[width=85mm]{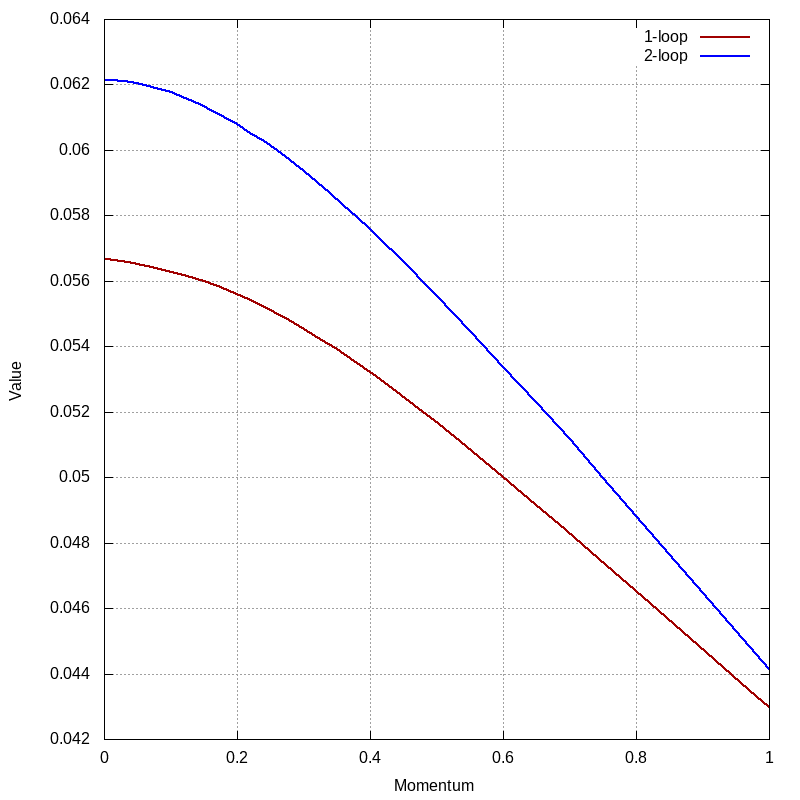}
\includegraphics[width=85mm]{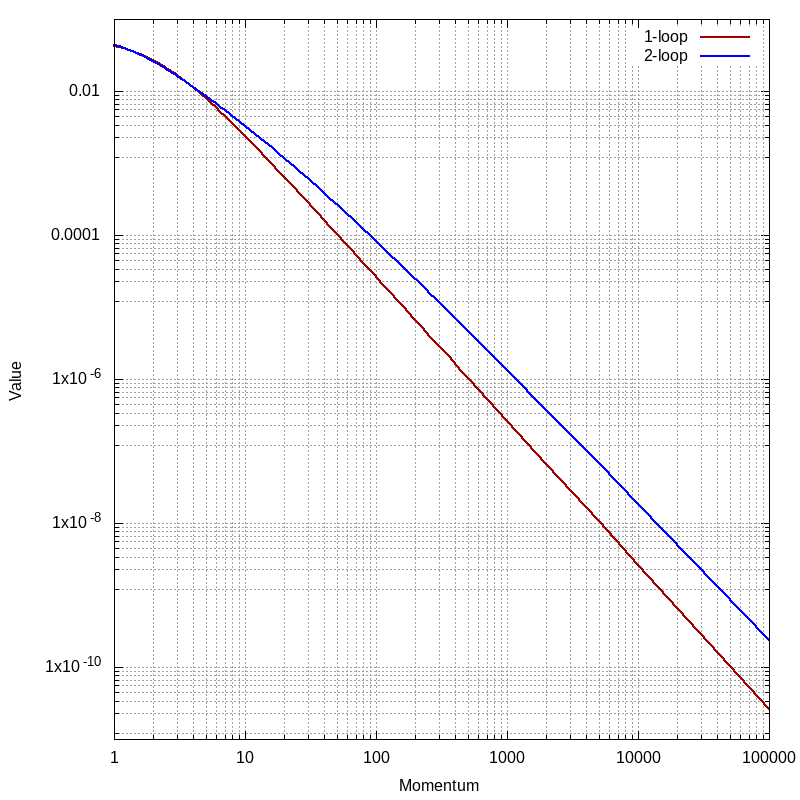}
\caption{The values of $\widetilde{V}_{13}(p)$ (1-loop) and $\widetilde{V}_{23}(p)$ (2-loop). 
The numerical data have been smoothed to enhance visual clarity.
\label{figgraphs_zalpha_3}
}
\end{figure*}

\begin{figure*}
\includegraphics[width=85mm]{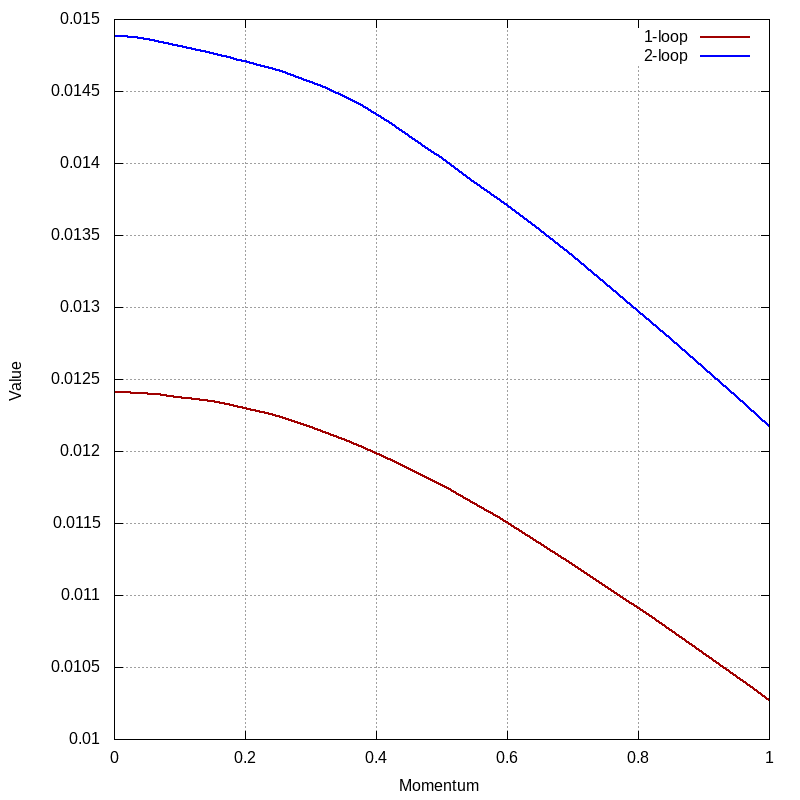}
\includegraphics[width=85mm]{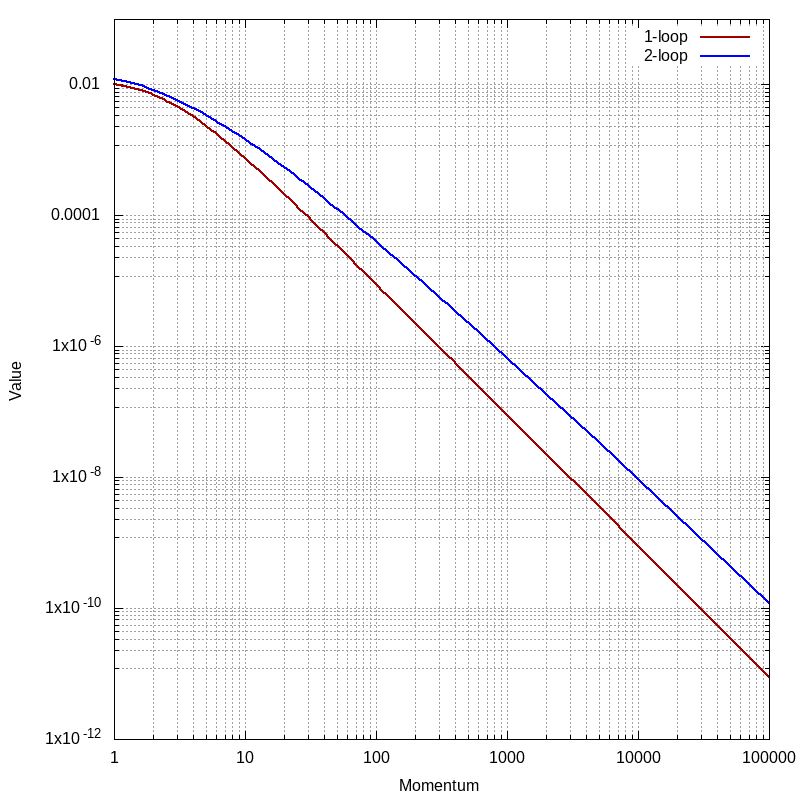}
\caption{The values of $\widetilde{V}_{15}(p)$ (1-loop) and $\widetilde{V}_{25}(p)$ (2-loop).
\label{figgraphs_zalpha_5}
}
\end{figure*}

\begin{figure*}
\includegraphics[width=85mm]{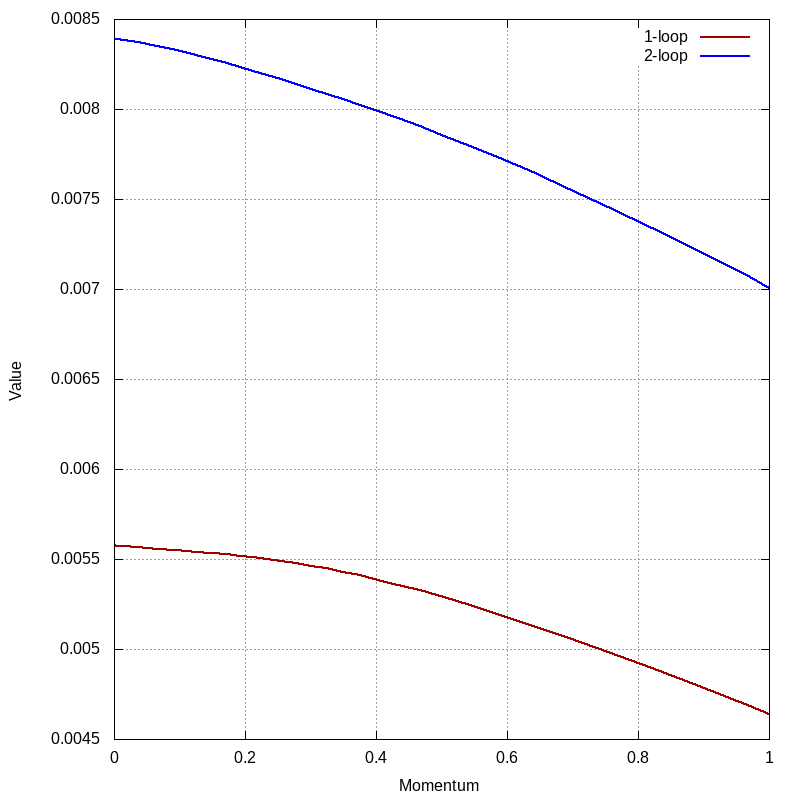}
\includegraphics[width=85mm]{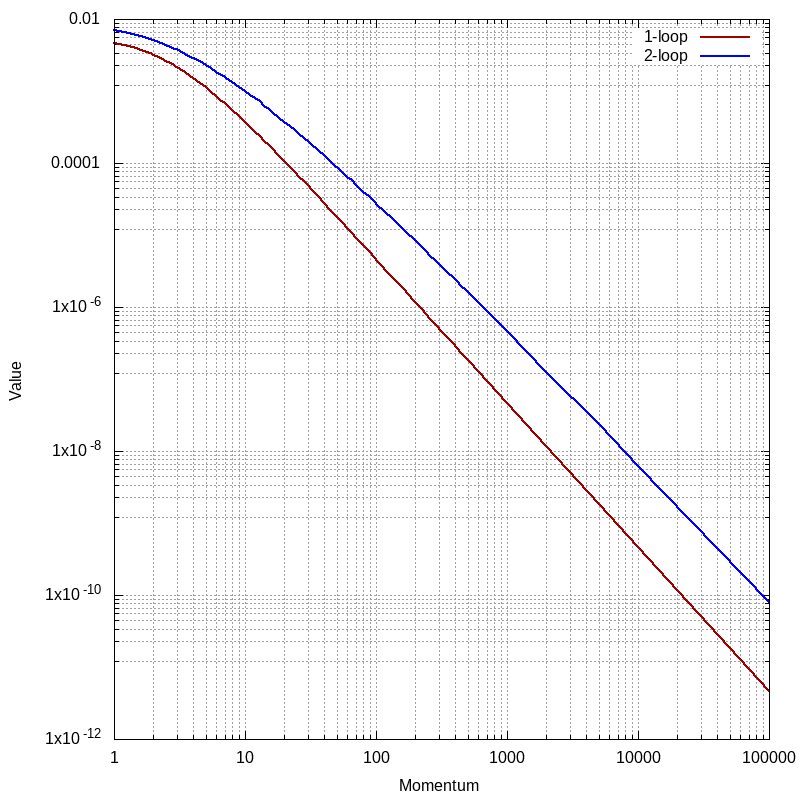}
\caption{The values of $\widetilde{V}_{17}(p)$ (1-loop) and $\widetilde{V}_{27}(p)$ (2-loop).
\label{figgraphs_zalpha_7}
}
\end{figure*}

Figs.~\ref{figgraphs_zalpha_3}, \ref{figgraphs_zalpha_5}, and \ref{figgraphs_zalpha_7} present plots of the two-loop potentials
$\widetilde{V}_{23}(\vsp{p})$, $\widetilde{V}_{25}(\vsp{p})$, and $\widetilde{V}_{27}(\vsp{p})$ in comparison with their one-loop counterparts,
$\widetilde{V}_{13}(\vsp{p})$, $\widetilde{V}_{15}(\vsp{p})$, and $\widetilde{V}_{17}(\vsp{p})$.
Note that the one- and two-loop prefactors were pulled out in the definition of $\widetilde{V}_{ij}$,
so the normalized one- and two-loop potentials are expected to be of the same order of magnitude.
We observe that the normalized two-loop potentials behave
similarly to their one-loop counterparts for small momenta, approaching constant values as $\vsp{p} \to \vspz$.
However, they behave differently at large momenta, with two-loop potentials decreasing more slowly than their
one-loop analogues.

We now examine the energy shifts induced by the two-loop VP potentials in hydrogen-like ions. It is convenient to
represent them in term of dimensionless function $G(\Za)$, pulling out their leading
$\alpha$, $\Za$, and $n$ dependence,
\begin{align}\label{eq:GZa}
\delta E = \left(\frac{\alpha}{\pi}\right)^2\,\frac{(\Za)^6}{n^3}\,G(\Za)\,,
\end{align}
where $n$ is the principal quantum number of the reference state.

Table~\ref{tab:en1} presents our numerical results for individual two-loop
VP contributions for the $1s$ state of hydrogen-like ions.
The columns labeled $\lbr V_{23}\rbr_{\rm pnt}$, $\lbr V_{25}\rbr_{\rm pnt}$,
and $\lbr V_{27}\rbr_{\rm pnt}$ represent
expectation values of the potentials $V_{23}$, $V_{25}$, and  $V_{27}$
evaluated with the point-nucleus
Dirac wave functions. 
The column labeled FNS
shows our estimations of the finite nuclear size (fns) correction,
$\lbr V_{23}+ V_{25}+V_{27}\rbr_{\rm fns}$.
It was obtained by multiplying the point-nucleus result $\lbr V_{23}+ V_{25}+V_{27}\rbr_{\rm pnt}$
by the relative fns correction for the $V_{21}$ expectation value \cite{yerokhin:08:twoloop},
$\lbr V_{21}\rbr_{\rm fns}/\lbr V_{21}\rbr_{\rm pnt}$, with uncertainty of 50\%.

The last column of the table presents the total results for the higher-order two-loop
VP correction,
\begin{align}
\delta E_{\rm VP23+} =&\ \lbr V_{23}\rbr_{\rm pnt}+ \lbr V_{25}\rbr_{\rm pnt}
      + \lbr V_{27}\rbr_{\rm pnt}
      \nonumber \\ &
       + \lbr V_{23}+ V_{25}+V_{27}\rbr_{\rm fns}\,.
\end{align}
The uncertainty due to omitted higher-order terms ($V_{29}$, etc.) is
not shown explicitly but included into the total uncertainty in the last column.
It was estimated by scaling the $\lbr V_{27}\rbr$ contribution
by the factor $2\,\lbr V_{27}\rbr/\lbr V_{25}\rbr$.

Our result for hydrogen $G(1\alpha) = 0.192$ is close to the analytical
$Z=0$ value obtained by
Krachkov and Lee $G(0\alpha) = 0.1954\ldots$ \cite{lee_krachkov_2023}.
In the high-$Z$ region, our results are consistent with previous estimates.
In particular, for the
$1s$ state of uranium, Ref.~\cite{yerokhin:08:twoloop} estimated
$G(92\alpha) = \pm 1.3$, whereas we now obtain
$G(92\alpha) = 0.169\,(4)$. It is remarkable that the calculated
correction turned out to be
an order of magnitude smaller than previously anticipated.

In Table~\ref{tab:vpvp}, we collect all two-loop VP contributions to transition energies of Li-like bismuth and uranium. The two leading terms -- those arising from the {\KS} potential $V_{21}$ and from
the second-order iteration of the one-loop VP potential $(V_1, V_1)$
-- were previously evaluated in Ref.~\cite{yerokhin:08:twoloop}. The remaining entries in the table represent higher-order two-loop VP corrections calculated in this work.

Table~\ref{tab:en2} summarizes our numerical results for the  higher-order two-loop
VP correction $\delta E_{\rm VP23+}$ for the $1s$, $2s$, $2p_{1/2}$, and $2p_{3/2}$ states of hydrogen-like
ions. These results extend calculations of two-loop corrections reported in
Ref.~\cite{yerokhin:08:twoloop} and complete the treatment of the two-loop
vacuum polarization effect.

We now turn to the experimental consequences of our calculations.
Table~\ref{tab:en3} summarises individual theoretical contributions
to the transition energies in Li-like
bismuth and uranium, for which accurate experimental results are available.
For the comprehensive summary of various theoretical corrections in Li-like
ions we refer the reader to the recent study \cite{yerokhin:25:Lilike:arxiv}.
As can be seen from Table~\ref{tab:en3},
after our calculation of the two-loop vacuum-polarization (VPVP) correction,
the main theoretical uncertainty for both bismuth and uranium
now arises from the last uncalculated two-loop QED
correction, namely, the self-energy with the VP insertion in the photon line (SVPE).

We observe that the updated theoretical predictions
of the transition energies have approximately twice smaller uncertainties
than the earlier results of Ref.~\cite{yerokhin:06:prl}. The
agreement with experiment is excellent in the case of bismuth.
For uranium, there is a slight tension between the theoretical and
experimental values (1.5$\sigma$ with the experiment
\cite{beiersdorfer:05}, 2.0$\sigma$ with that of Ref.~\cite{brandau:04},
and 1.4$\sigma$ with Ref.~\cite{schweppe:91}).
A possible explanation of this tension is an underestimated uncertainty
in the currently accepted nuclear charge radius of $^{238}$U as listed in the tabulation
\cite{angeli:13}.
This interpretation is consistent with the recent criticism by Ohayon
\cite{ohayon:25:radii}, who argued that the model dependence of the nuclear charge
distribution had not been properly accounted for in Ref.~\cite{angeli:13}.
A similar issue was recently identified
for the doubly-magic
nucleus $^{208}$Pb, where a state-of-the-art
re-analysis of historical muonic spectroscopy data led to a 3$\sigma$ shift in the accepted nuclear radius \cite{sun:25}.

\section{Summary}

We performed calculations of Coulomb corrections to the leading-order
two-loop vacuum-polarization potential derived 70 years ago
by G. K{\"a}ll{\'e}n and A. Sabry \cite{kaellen:55}. Specifically,
we derived the two-loop vacuum-polarization potentials
$V_{23}$, $V_{25}$, and $V_{27}$ that
are induced by Feynman diagrams with three, five, and seven Coulomb interactions inside the vacuum-polarization loop,
respectively.
The potentials were computed in momentum representation and stored
on a grid. Furthermore, we
evaluated expectation values of these potentials with the reference-state wave function
and obtained the corresponding energy shifts
for H-like ions in a wide range of nuclear charges $Z$.
In the limit of small $Z$, our results agree with the
analytical result of Krachkov and Lee \cite{lee_krachkov_2023}.

Our calculation eliminates one of the two largest uncertainties in
theoretical values of the two-loop Lamb shift and represents an important
step towards completing the long-standing project of
calculating of the full set of one-electron two-loop QED
effects. 

Comparison of the
updated theoretical predictions for the $2p_j$-$2s$ transition
energies in bismuth and uranium
with available experiments results yields some of the best
tests of bound-state QED theory in the strong nuclear binding
field. We find excellent agreement in the case of bismuth but a small
tension of $1.4$ - $2.0\sigma$ for uranium, which may stem from
the insufficient knowledge of the nuclear charge radius.

In the future, it might be worthwhile to obtain the potentials $V_{2j}$ in the coordinate representation, which is much more
convenient for practical applications than the momentum one. This could be achieved through a numerical Fourier transformation,
but it requires further knowledge of the behaviour of the potentials near $\vsp{p}=0$ and $\vsp{p}=\infty$.


%
%
%
\ \\ \ \\

\begin{table}[htb]
\begin{center}
\caption{
Energy shifts induced by the two-loop vacuum-polarization potentials
$V_{23}$, $V_{25}$, and $V_{27}$, for the $1s$ state
of hydrogen-like ions, in terms of the function $G(\Za)$ defined by Eq.~(\ref{eq:GZa}).
Expectation values of the potentials are evaluated for the point nuclear model;
FNS denotes estimates of the finite nuclear size correction.
\label{tab:en1}
}
\begin{ruledtabular}
\begin{tabular}{lw{1.7}w{1.6}w{1.6}w{3.6}w{1.6}}
           \multicolumn{1}{c}{$Z$}
                &  \multicolumn{1}{c}{$\lbr V_{23}\rbr_{\rm pnt}$}
                    &  \multicolumn{1}{c}{$\lbr V_{25}\rbr_{\rm pnt} $ }
                            &  \multicolumn{1}{c}{$\lbr V_{27}\rbr_{\rm pnt} $ }
                        &  \multicolumn{1}{c}{FNS}
                                &  \multicolumn{1}{c}{Total}
\\ \hline\\[-9pt]
%
%
  1   &   0.1915\,(7)      &                        &                         &                          &        0.1915\,(7) \\
  2   &   0.1881\,(5)      &                        &                         &                          &        0.1881\,(5) \\
  3   &   0.1849\,(3)      &                        &                         &                          &        0.1849\,(3) \\
  5   &   0.1789\,(2)      &        0.0001          &                         &                          &        0.1789\,(2) \\
  7   &   0.1734\,(1)      &        0.0001          &                         &                          &        0.1735\,(1) \\
 10   &   0.1660\,(1)      &        0.0002          &                         &                          &        0.1662\,(1) \\
 15   &   0.1557           &        0.0005          &                         &                          &        0.1562 \\
 20   &   0.1473           &        0.0008          &                         &                          &        0.1481 \\
 25   &   0.1405           &        0.0013          &                         &                          &        0.1417 \\
 30   &   0.1349           &        0.0018          &         0.0001          &         -0.0001          &        0.1366 \\
 35   &   0.1304           &        0.0024          &         0.0001          &         -0.0001\,(1)     &        0.1328\,(1) \\
 40   &   0.1269           &        0.0031          &         0.0002          &         -0.0002\,(1)     &        0.1300\,(1) \\
 45   &   0.1243           &        0.0039          &         0.0003          &         -0.0002\,(1)     &        0.1283\,(1) \\
 50   &   0.1226\,(1)      &        0.0049          &         0.0004          &         -0.0003\,(2)     &        0.1275\,(2) \\
 55   &   0.1216\,(1)      &        0.0059          &         0.0006          &         -0.0005\,(2)     &        0.1277\,(3) \\
 60   &   0.1215\,(1)      &        0.0072          &         0.0008          &         -0.0007\,(3)     &        0.1289\,(4) \\
 65   &   0.1223\,(1)      &        0.0087          &         0.0012          &         -0.0010\,(5)     &        0.1311\,(6) \\
 70   &   0.1240\,(1)      &        0.0104          &         0.0017          &         -0.0015\,(7)     &        0.1346\,(9) \\
 75   &   0.1267\,(1)      &        0.0125          &         0.0023\,(1)     &         -0.002\,(1)      &        0.139\,(1) \\
 80   &   0.1308\,(1)      &        0.0149          &         0.0032\,(1)     &         -0.003\,(2)      &        0.146\,(2) \\
 83   &   0.1339\,(1)      &        0.0167          &         0.0039\,(1)     &         -0.004\,(2)      &        0.151\,(3) \\
 90   &   0.1439\,(2)      &        0.0217          &         0.0060\,(2)     &         -0.007\,(4)      &        0.165\,(5) \\
 92   &   0.1476\,(2)      &        0.0235\,(1)     &         0.0068\,(2)     &         -0.009\,(4)      &        0.169\,(6) \\
100   &   0.1677\,(3)      &        0.0328\,(1)     &         0.0114\,(3)     &         -0.017\,(9)      &        0.195\,(12) \\
\end{tabular}
\end{ruledtabular}
\end{center}
\end{table}

%
%
\begin{table}[htb]
\begin{center}
\caption{
Individual two-loop vacuum-polarization
contributions to transition energies in Li-like bismuth and uranium, in eV.
Subscripts ``ext'' and ``pnt'' indicate results obtained for the extended-size and the point
nuclear models; ``fns'' denotes the finite nuclear size correction.
Uncertainties due to nuclear charge radii and models are now shown.
\label{tab:vpvp}
}
\begin{ruledtabular}
\begin{tabular}{lw{2.9}w{2.9}}
 Term
              &  \multicolumn{1}{c}{$2p_{3/2}$-$2s$, $Z=83$}
                &  \multicolumn{1}{c}{$2p_{1/2}$-$2s$, $Z=92$}
\\ \hline\\[-9pt]
$\lbr V_{21} \rbr_{\rm ext}$         &  0.06718       &  0.10111       \\
$\lbr V_{1},V_{1}\rbr_{\rm ext}   $  &  0.01573       &  0.03501       \\
$\lbr V_{23}  \rbr_{\rm pnt}$        & -0.00276       & -0.00474\,(1)  \\
$\lbr V_{25}  \rbr_{\rm pnt}$        & -0.00036       & -0.00082       \\
$\lbr V_{27}  \rbr_{\rm pnt}$        & -0.00009       & -0.00025\,(1)  \\
$\lbr V_{23+} \rbr_{\rm fns}$        &  0.00009\,(5)  &  0.00034\,(20)    \\
Sum                                  &  0.07979\,(6)  &  0.13066\,(27)   \\
Ref.~\cite{yerokhin:06:prl}
                                     &  0.083\,(25)   &  0.136\,(46)   \\
\end{tabular}
\end{ruledtabular}
\end{center}
\end{table}

%
%
%
\begin{table}[htb]
\begin{center}
\caption{
The higher-order two-loop vacuum-polarization correction
$\delta E_{\rm VP23+}$
for different states
of hydrogen-like ions, in terms of the function $G(\Za)$ defined by Eq.~(\ref{eq:GZa}).
\label{tab:en2}
}
\begin{ruledtabular}
\begin{tabular}{lw{1.7}w{2.7}w{2.7}w{2.7}w{2.7}}
           \multicolumn{1}{c}{$Z$}
                &  \multicolumn{1}{c}{$1s$}
                    &  \multicolumn{1}{c}{$2s$ }
                        &  \multicolumn{1}{c}{$2p_{1/2}$ }
                            &  \multicolumn{1}{c}{$2p_{3/2}$ }
\\ \hline\\[-9pt]
%
%
  1    & 0.1915\,(7)            & 0.191\,(2)             & \\
  2    & 0.1881\,(5)            & 0.1882\,(8)            & \\
  3    & 0.1849\,(3)            & 0.1850\,(4)            & \\
  5    & 0.1789\,(2)            & 0.1789\,(3)            & 0.0001                  \\
  7    & 0.1735\,(1)            & 0.1735\,(1)            & 0.0002                 & 0.0001 \\
 10    & 0.1662\,(1)            & 0.1664                 & 0.0004                 & 0.0002 \\
 15    & 0.1562                 & 0.1567                 & 0.0009                 & 0.0004 \\
 20    & 0.1481                 & 0.1491                 & 0.0015                 & 0.0007 \\
 25    & 0.1417                 & 0.1434                 & 0.0022                 & 0.0010 \\
 30    & 0.1366                 & 0.1392                 & 0.0031                 & 0.0013 \\
 35    & 0.1328\,(1)            & 0.1364\,(1)            & 0.0042                 & 0.0016 \\
 40    & 0.1300\,(1)            & 0.1350\,(1)            & 0.0054                 & 0.0019 \\
 45    & 0.1283\,(1)            & 0.1348\,(1)            & 0.0069                 & 0.0023 \\
 50    & 0.1275\,(2)            & 0.1359\,(2)            & 0.0087                 & 0.0026 \\
 55    & 0.1277\,(3)            & 0.1384\,(3)            & 0.0109                 & 0.0030 \\
 60    & 0.1289\,(4)            & 0.1425\,(5)            & 0.0134                 & 0.0034 \\
 65    & 0.1311\,(6)            & 0.1482\,(7)            & 0.0166                 & 0.0038 \\
 70    & 0.1346\,(9)            & 0.156\,(1)             & 0.0204\,(1)            & 0.0042 \\
 75    & 0.139\,(1)             & 0.166\,(2)             & 0.0252\,(1)            & 0.0046 \\
 80    & 0.146\,(2)             & 0.179\,(3)             & 0.0313\,(2)            & 0.0050 \\
 83    & 0.151\,(3)             & 0.189\,(4)             & 0.0357\,(4)            & 0.0053 \\
 90    & 0.165\,(5)             & 0.217\,(7)             & 0.049\,(1)             & 0.0060 \\
 92    & 0.169\,(6)             & 0.227\,(8)             & 0.054\,(1)             & 0.0062 \\
100    & 0.195\,(12)            & 0.28\,(2)              & 0.081\,(4)             & 0.0071\,(1) \\
\end{tabular}
\end{ruledtabular}
\end{center}
\end{table}

%
%
\begin{table*}[t]
\caption{Theoretical contributions to transition energies in Li-like
bismuth and uranium, in eV. All theory except two-loop vacuum-polarization (VPVP) is from
Ref.~\cite{yerokhin:25:Lilike:arxiv}.
When two uncertainties are specified, the first is the purely theoretical
uncertainty, whereas the second is due to the nuclear radius.
The nuclear charge radii used are $5.5211\,(26)\,$fm for bismuth and $5.8571\,(33)\,$fm for uranium
\cite{angeli:13}.
 \label{tab:en3} }
\begin{ruledtabular}
\begin{tabular}{llw{5.7}w{5.7}c}
         && \multicolumn{1}{c}{$2p_{3/2}$-$2s$, $^{209}$Bi$^{80+}$}
                              & \multicolumn{1}{c}{$2p_{1/2}$-$2s$, $^{238}$U$^{89+}$}
                              & Ref.
                                  \\
 \hline\\[-9pt]
Structure            &&   2814.395\,(3)(26)&   322.286\,(5)(34) \\
One-loop QED         &&    -27.486\,(1)    &   -42.929\,(1)     \\
Screened one-loop QED&&      1.140\,(3)    &     1.193\,(16)    \\
Two-loop QED  & SESE  &      0.141         &     0.296          \\
              & SEVP  &     -0.095         &    -0.188          \\
              & SVPE  &     -0.002\,(31)   &    -0.004\,(66)    \\
              & VPVP  &      0.080         &     0.131      & This work\\
Screened two-loop QED&&     -0.007\,(5)    &    -0.012\,(9)    \\
Nuclear recoil       &&     -0.062\,(2)    &    -0.063\,(3)    \\
Nuclear polarization \& deformation   &&      0.013\,(13)   &     0.058\,(20)   \\
 \hline\\[-9pt]
Theory               &&   2788.116\,(35)(26)& 280.767\,(72)(34)    \\
Previous theory      &&   2788.12\,(7)     &  280.76 \,(14)  & \cite{yerokhin:06:prl}\\
Experiment           &&   2788.14\,(4)\,
                                            & 280.645\,(15)\, &  \cite{beiersdorfer:98,beiersdorfer:05} \\
                     &&                     & 280.516\,(99)\, &  \cite{brandau:04} \\
                     &&                     & 280.59\,(10)\,  & \cite{schweppe:91} \\
\end{tabular}
\end{ruledtabular}
\end{table*}

\ \\ 

\bibliography{hfst,all_phys}

\begin{widetext}

\begin{center}
{\large\bf SUPPLEMENTAL MATERIALS}
\end{center}

\begin{center}
{\bf The values of $\widetilde{V}_{13}(\vsp{p})$ for different $|\vsp{p}|$. \\
The format is:\\
$|\vsp{p}|$ Value Error}
\end{center}
\begin{lstlisting}
0.001 5.666592e-02 1.927567e-05
0.003 5.667512e-02 2.123566e-05
0.005 5.668134e-02 2.202271e-05
0.01 5.665221e-02 2.282921e-05
0.03 5.663722e-02 2.366989e-05
0.05 5.658595e-02 2.390767e-05
0.1 5.635320e-02 2.399248e-05
0.15 5.606263e-02 2.386761e-05
0.16 5.602655e-02 2.386898e-05
0.17 5.588753e-02 2.381936e-05
0.18 5.582369e-02 2.371460e-05
0.19 5.576325e-02 2.363793e-05
0.2 5.564468e-02 2.359762e-05
0.22 5.543281e-02 2.350252e-05
0.24 5.527522e-02 2.340596e-05
0.26 5.507876e-02 2.328949e-05
0.28 5.484517e-02 2.312935e-05
0.3 5.463553e-02 2.291152e-05
0.32 5.434417e-02 2.275463e-05
0.34 5.411447e-02 2.251352e-05
0.36 5.384436e-02 2.233527e-05
0.38 5.357801e-02 2.221943e-05
0.4 5.324917e-02 2.198082e-05
0.42 5.302752e-02 2.179395e-05
0.44 5.271470e-02 2.166618e-05
0.46 5.244488e-02 2.114638e-05
0.48 5.213428e-02 2.111238e-05
0.5 5.179877e-02 2.090170e-05
0.55 5.101796e-02 2.056741e-05
0.6 5.016132e-02 1.992483e-05
0.65 4.930687e-02 1.948309e-05
0.7 4.840000e-02 1.859623e-05
0.75 4.753753e-02 1.845565e-05
0.8 4.660152e-02 1.779020e-05
0.85 4.567025e-02 1.721498e-05
0.9 4.473625e-02 1.579775e-05
0.95 4.380516e-02 1.564522e-05
1 4.284435e-02 1.598265e-05
1.1 4.097388e-02 1.474686e-05
1.2 3.916924e-02 1.378217e-05
1.3 3.736380e-02 1.231272e-05
1.4 3.562180e-02 1.243989e-05
1.5 3.398351e-02 1.201967e-05
1.6 3.236674e-02 1.102733e-05
1.7 3.083669e-02 1.069486e-05
1.8 2.937064e-02 9.841465e-06
1.9 2.797224e-02 9.569615e-06
2 2.665561e-02 8.829822e-06
2.1 2.540616e-02 8.608763e-06
2.2 2.421483e-02 7.762967e-06
2.3 2.309070e-02 7.781969e-06
2.4 2.204677e-02 7.625355e-06
2.5 2.104391e-02 6.705683e-06
2.6 2.010459e-02 6.398186e-06
2.7 1.922024e-02 6.112939e-06
2.8 1.836845e-02 6.337615e-06
2.9 1.758590e-02 6.067025e-06
3 1.684927e-02 5.499073e-06
3.5 1.368058e-02 4.616588e-06
4 1.128446e-02 3.844855e-06
4.5 9.427023e-03 3.254007e-06
5 7.981260e-03 2.649901e-06
5.5 6.837369e-03 2.360781e-06
6 5.913599e-03 2.019058e-06
6.5 5.156787e-03 1.704806e-06
7 4.537576e-03 1.521206e-06
7.5 4.019345e-03 1.368414e-06
8 3.585191e-03 1.235956e-06
9 2.897397e-03 1.003888e-06
10 2.390539e-03 8.491202e-07
11 2.001483e-03 7.134627e-07
12 1.701177e-03 6.075101e-07
13 1.463280e-03 5.231935e-07
14 1.269712e-03 4.550238e-07
15 1.113245e-03 3.991955e-07
16 9.835133e-04 3.589900e-07
18 7.834574e-04 2.855518e-07
20 6.385828e-04 2.321464e-07
22 5.298523e-04 1.950021e-07
24 4.469950e-04 1.636867e-07
26 3.821538e-04 1.391772e-07
28 3.301306e-04 1.211985e-07
30 2.881185e-04 1.051275e-07
32 2.537100e-04 9.307447e-08
36 2.008662e-04 7.437728e-08
40 1.630168e-04 6.073053e-08
44 1.348979e-04 4.993811e-08
48 1.136044e-04 4.212976e-08
52 9.683167e-05 3.598155e-08
56 8.351817e-05 3.106340e-08
60 7.277332e-05 2.730919e-08
64 6.400199e-05 2.377425e-08
72 5.065598e-05 1.904116e-08
80 4.102661e-05 1.533675e-08
88 3.391120e-05 1.268408e-08
96 2.850523e-05 1.065111e-08
104 2.430117e-05 9.121775e-09
112 2.094956e-05 7.843366e-09
120 1.825083e-05 6.853557e-09
128 1.604567e-05 6.035124e-09
144 1.268733e-05 4.778215e-09
160 1.027302e-05 3.870110e-09
176 8.493665e-06 3.194107e-09
192 7.134647e-06 2.689955e-09
208 6.082945e-06 2.283253e-09
224 5.240661e-06 1.969491e-09
240 4.564501e-06 1.715073e-09
256 4.015411e-06 1.498947e-09
288 3.171708e-06 1.190925e-09
320 2.569565e-06 9.679213e-10
352 2.123570e-06 7.979248e-10
384 1.785173e-06 6.735130e-10
416 1.520039e-06 5.734810e-10
448 1.309756e-06 4.938253e-10
480 1.141887e-06 4.307909e-10
512 1.003316e-06 3.775895e-10
1000 2.632820e-07 9.942789e-11
2000 6.575227e-08 2.485423e-11
5000 1.051752e-08 3.994115e-12
10000 2.630485e-09 9.999783e-13
100000 2.632339e-11 1.004318e-14
\end{lstlisting}

\begin{center}
{\bf The values of $\widetilde{V}_{15}(\vsp{p})$ for different $|\vsp{p}|$. \\
The format is:\\
$|\vsp{p}|$ Value Error}
\end{center}
\begin{lstlisting}
0.001 1.241017e-02 3.838217e-06
0.003 1.241453e-02 4.495686e-06
0.005 1.241073e-02 4.827981e-06
0.01 1.241216e-02 5.247710e-06
0.03 1.241353e-02 5.903178e-06
0.05 1.240379e-02 6.222347e-06
0.1 1.238242e-02 6.585745e-06
0.15 1.235380e-02 6.872912e-06
0.16 1.236061e-02 6.862367e-06
0.17 1.232566e-02 6.847634e-06
0.18 1.232862e-02 6.837045e-06
0.19 1.231664e-02 6.320527e-06
0.2 1.230846e-02 6.811642e-06
0.22 1.228285e-02 6.606248e-06
0.24 1.225994e-02 6.958149e-06
0.26 1.223180e-02 6.931245e-06
0.28 1.220449e-02 6.910712e-06
0.3 1.218801e-02 6.880540e-06
0.32 1.215601e-02 6.854154e-06
0.34 1.212263e-02 6.635587e-06
0.36 1.207927e-02 7.008458e-06
0.38 1.204072e-02 6.980752e-06
0.4 1.200030e-02 6.556387e-06
0.42 1.197486e-02 6.924649e-06
0.44 1.193008e-02 6.894249e-06
0.46 1.189075e-02 6.873700e-06
0.48 1.186298e-02 6.837655e-06
0.5 1.178929e-02 6.806703e-06
0.55 1.167588e-02 6.531692e-06
0.6 1.154784e-02 6.652985e-06
0.65 1.141694e-02 5.892314e-06
0.7 1.127124e-02 6.492984e-06
0.75 1.112409e-02 6.409372e-06
0.8 1.097235e-02 5.988111e-06
0.85 1.080950e-02 6.237593e-06
0.9 1.063887e-02 6.147893e-06
0.95 1.047922e-02 5.879373e-06
1 1.031686e-02 5.970818e-06
1.1 9.958946e-03 5.433422e-06
1.2 9.607788e-03 5.264254e-06
1.3 9.261166e-03 5.092235e-06
1.4 8.916827e-03 4.924353e-06
1.5 8.573316e-03 4.757701e-06
1.6 8.237361e-03 4.593211e-06
1.7 7.912279e-03 4.432706e-06
1.8 7.591409e-03 4.277177e-06
1.9 7.279726e-03 4.023000e-06
2 6.986904e-03 3.880366e-06
2.1 6.700069e-03 3.745260e-06
2.2 6.435919e-03 3.613143e-06
2.3 6.171647e-03 3.486711e-06
2.4 5.926459e-03 3.365861e-06
2.5 5.682185e-03 3.171444e-06
2.6 5.467670e-03 3.063629e-06
2.7 5.242747e-03 2.958614e-06
2.8 5.043307e-03 2.858763e-06
2.9 4.851190e-03 2.764642e-06
3 4.665623e-03 2.673693e-06
3.5 3.868133e-03 2.223330e-06
4 3.243598e-03 1.869566e-06
4.5 2.751097e-03 1.589228e-06
5 2.359335e-03 1.364744e-06
5.5 2.040269e-03 1.182913e-06
6 1.782375e-03 1.053158e-06
6.5 1.568749e-03 9.264490e-07
7 1.390251e-03 8.206695e-07
7.5 1.239770e-03 7.314274e-07
8 1.112194e-03 6.348622e-07
9 9.093719e-04 5.266605e-07
10 7.560552e-04 4.499673e-07
11 6.385482e-04 3.838764e-07
12 5.454940e-04 3.311725e-07
13 4.714319e-04 2.846122e-07
14 4.116890e-04 2.469492e-07
15 3.618358e-04 2.187242e-07
16 3.212405e-04 1.927438e-07
18 2.573253e-04 1.577581e-07
20 2.107492e-04 1.286838e-07
22 1.755640e-04 1.079590e-07
24 1.483566e-04 9.164509e-08
26 1.271408e-04 7.800504e-08
28 1.101781e-04 6.711040e-08
30 9.631769e-05 5.929465e-08
32 8.489545e-05 5.277380e-08
36 6.743859e-05 4.186778e-08
40 5.481014e-05 3.395778e-08
44 4.544860e-05 2.824207e-08
48 3.825352e-05 2.382971e-08
52 3.265862e-05 2.021996e-08
56 2.821551e-05 1.757191e-08
60 2.464243e-05 1.540153e-08
64 2.165005e-05 1.352251e-08
72 1.713994e-05 1.070436e-08
80 1.389594e-05 8.712171e-09
88 1.149907e-05 7.217871e-09
96 9.672703e-06 6.070326e-09
104 8.239352e-06 5.170165e-09
112 7.107573e-06 4.451638e-09
120 6.187294e-06 3.886821e-09
128 5.449932e-06 3.421130e-09
144 4.308554e-06 2.705100e-09
160 3.489833e-06 2.188097e-09
176 2.884494e-06 1.809968e-09
192 2.426052e-06 1.525930e-09
208 2.066929e-06 1.298052e-09
224 1.781821e-06 1.120419e-09
240 1.552982e-06 9.761883e-10
256 1.365643e-06 8.601957e-10
288 1.078737e-06 6.777213e-10
320 8.738880e-07 5.497949e-10
352 7.220553e-07 4.543698e-10
384 6.064191e-07 3.813159e-10
416 5.170439e-07 3.251049e-10
448 4.452672e-07 2.802892e-10
480 3.883188e-07 2.445021e-10
512 3.411115e-07 2.145281e-10
1000 8.959559e-08 5.638504e-11
2000 2.236214e-08 1.405666e-11
5000 3.582579e-09 2.250331e-12
10000 8.955352e-10 5.619415e-13
100000 8.955613e-12 5.600730e-15
\end{lstlisting}

\begin{center}
{\bf The values of $\widetilde{V}_{17}(\vsp{p})$ for different $|\vsp{p}|$. \\
The format is:\\
$|\vsp{p}|$ Value Error}
\end{center}
\begin{lstlisting}
0.001 5.557598e-03 1.057891e-05
0.003 5.578052e-03 1.099507e-05
0.005 5.558203e-03 1.029535e-05
0.01 5.576865e-03 1.100024e-05
0.03 5.580508e-03 1.098539e-05
0.05 5.548063e-03 1.061880e-05
0.1 5.549179e-03 1.098615e-05
0.15 5.524537e-03 1.055336e-05
0.16 5.540696e-03 1.093785e-05
0.17 5.537790e-03 1.085025e-05
0.18 5.517274e-03 1.092717e-05
0.19 5.516223e-03 1.093344e-05
0.2 5.512599e-03 1.052011e-05
0.22 5.514389e-03 1.090634e-05
0.24 5.503903e-03 1.050591e-05
0.26 5.486184e-03 1.086847e-05
0.28 5.465306e-03 1.047395e-05
0.3 5.486709e-03 1.036473e-05
0.32 5.474305e-03 1.043258e-05
0.34 5.421021e-03 1.082086e-05
0.36 5.448482e-03 1.037893e-05
0.38 5.412192e-03 1.074651e-05
0.4 5.393981e-03 1.072203e-05
0.42 5.377292e-03 1.071513e-05
0.44 5.354546e-03 1.028539e-05
0.46 5.367959e-03 1.062183e-05
0.48 5.319450e-03 1.023345e-05
0.5 5.300289e-03 1.057416e-05
0.55 5.246022e-03 1.050086e-05
0.6 5.203898e-03 1.038422e-05
0.65 5.136437e-03 9.927112e-06
0.7 5.068256e-03 1.021197e-05
0.75 5.027404e-03 1.011497e-05
0.8 4.956233e-03 9.648953e-06
0.85 4.871527e-03 1.002499e-05
0.9 4.808983e-03 9.788607e-06
0.95 4.724051e-03 9.666710e-06
1 4.650442e-03 9.538085e-06
1.1 4.514378e-03 8.955080e-06
1.2 4.358656e-03 8.422036e-06
1.3 4.216017e-03 8.461582e-06
1.4 4.052566e-03 8.207144e-06
1.5 3.897960e-03 7.829192e-06
1.6 3.745086e-03 7.460895e-06
1.7 3.609283e-03 7.222010e-06
1.8 3.479433e-03 7.010789e-06
1.9 3.323587e-03 6.550159e-06
2 3.209635e-03 6.775356e-06
2.1 3.069573e-03 6.348815e-06
2.2 2.961799e-03 6.128629e-06
2.3 2.842006e-03 5.768738e-06
2.4 2.728850e-03 5.741704e-06
2.5 2.616025e-03 5.224752e-06
2.6 2.528113e-03 5.376464e-06
2.7 2.433822e-03 5.111279e-06
2.8 2.334495e-03 4.747354e-06
2.9 2.249436e-03 4.598571e-06
3 2.167114e-03 4.459743e-06
3.5 1.808271e-03 3.728170e-06
4 1.523992e-03 3.227748e-06
4.5 1.306768e-03 2.676170e-06
5 1.121196e-03 2.361114e-06
5.5 9.748550e-04 2.146369e-06
6 8.560427e-04 1.831960e-06
6.5 7.537169e-04 1.611704e-06
7 6.731796e-04 1.458005e-06
7.5 6.008734e-04 1.297509e-06
8 5.418029e-04 1.161650e-06
9 4.432576e-04 9.626158e-07
10 3.707409e-04 8.105123e-07
11 3.144633e-04 6.848208e-07
12 2.697436e-04 5.726206e-07
13 2.334134e-04 5.015360e-07
14 2.041114e-04 4.553241e-07
15 1.811205e-04 3.968850e-07
16 1.605401e-04 3.477409e-07
18 1.291806e-04 2.886125e-07
20 1.060465e-04 2.345766e-07
22 8.867662e-05 1.986292e-07
24 7.522425e-05 1.662705e-07
26 6.462705e-05 1.455315e-07
28 5.576901e-05 1.260220e-07
30 4.898271e-05 1.099926e-07
32 4.333879e-05 9.794867e-08
36 3.452278e-05 7.795246e-08
40 2.816651e-05 6.400693e-08
44 2.331138e-05 5.297404e-08
48 1.967708e-05 4.455826e-08
52 1.677935e-05 3.822107e-08
56 1.456628e-05 3.314079e-08
60 1.263078e-05 2.876513e-08
64 1.117314e-05 2.533629e-08
72 8.835826e-06 2.008837e-08
80 7.160059e-06 1.630498e-08
88 5.952645e-06 1.361157e-08
96 4.994122e-06 1.140366e-08
104 4.281996e-06 9.845486e-09
112 3.685126e-06 8.446458e-09
120 3.211282e-06 7.394413e-09
128 2.825505e-06 6.455079e-09
144 2.227741e-06 5.125879e-09
160 1.807852e-06 4.167447e-09
176 1.491499e-06 3.419543e-09
192 1.256132e-06 2.894183e-09
208 1.072305e-06 2.465260e-09
224 9.241834e-07 2.138811e-09
240 8.002070e-07 1.845520e-09
256 7.013934e-07 1.615510e-09
288 5.557832e-07 1.286736e-09
320 4.509016e-07 1.040620e-09
352 3.736795e-07 8.618912e-10
384 3.140434e-07 7.263655e-10
416 2.679088e-07 6.180440e-10
448 2.305899e-07 5.343662e-10
480 2.012440e-07 4.649190e-10
512 1.771967e-07 4.111226e-10
1000 4.628375e-08 1.072535e-10
2000 1.160432e-08 2.692646e-11
5000 1.857432e-09 4.323788e-12
10000 4.625948e-10 1.077523e-12
100000 4.609411e-12 1.075526e-14
\end{lstlisting}

\begin{center}
{\bf The values of $\widetilde{V}_{21}(\vsp{p})$ for different $|\vsp{p}|$. \\
The format is:\\
$|\vsp{p}|$ Value Error}
\end{center}
\begin{lstlisting}
0.001 -3.223444e-01 5.545325e-05
0.003 -3.222213e-01 4.173008e-05
0.005 -3.223851e-01 7.710039e-05
0.01 -3.222715e-01 5.545051e-05
0.03 -3.221460e-01 4.974589e-05
0.05 -3.222028e-01 6.362662e-05
0.1 -3.218325e-01 6.358478e-05
0.15 -3.212246e-01 3.849638e-05
0.16 -3.210057e-01 5.896813e-05
0.17 -3.208172e-01 4.679875e-05
0.18 -3.207175e-01 5.953653e-05
0.19 -3.204681e-01 6.335880e-05
0.2 -3.204130e-01 5.070822e-05
0.22 -3.199851e-01 6.053889e-05
0.24 -3.195341e-01 4.436070e-05
0.26 -3.190439e-01 5.865946e-05
0.28 -3.185994e-01 3.543599e-05
0.3 -3.181370e-01 7.760352e-05
0.32 -3.175335e-01 5.477009e-05
0.34 -3.169319e-01 5.155587e-05
0.36 -3.162656e-01 5.457785e-05
0.38 -3.157365e-01 7.925325e-05
0.4 -3.149313e-01 6.244747e-05
0.42 -3.141252e-01 5.428889e-05
0.44 -3.132787e-01 5.838525e-05
0.46 -3.125690e-01 7.647635e-05
0.48 -3.119299e-01 5.819083e-05
0.5 -3.108736e-01 5.382773e-05
0.55 -3.087507e-01 4.771850e-05
0.6 -3.062590e-01 5.672787e-05
0.65 -3.037136e-01 5.692811e-05
0.7 -3.010343e-01 6.016755e-05
0.75 -2.982096e-01 4.902926e-05
0.8 -2.952311e-01 3.882097e-05
0.85 -2.920998e-01 4.820609e-05
0.9 -2.888572e-01 4.776148e-05
0.95 -2.855715e-01 5.512555e-05
1 -2.822812e-01 5.303274e-05
1.1 -2.754320e-01 5.592513e-05
1.2 -2.683356e-01 3.911640e-05
1.3 -2.611184e-01 5.119926e-05
1.4 -2.538771e-01 6.442949e-05
1.5 -2.466575e-01 5.107722e-05
1.6 -2.395274e-01 3.679869e-05
1.7 -2.324017e-01 3.474820e-05
1.8 -2.253612e-01 3.326664e-05
1.9 -2.185100e-01 4.419295e-05
2 -2.118119e-01 4.506205e-05
2.1 -2.053687e-01 3.512490e-05
2.2 -1.990748e-01 3.373646e-05
2.3 -1.929593e-01 4.173574e-05
2.4 -1.870373e-01 5.142432e-05
2.5 -1.813327e-01 3.965400e-05
2.6 -1.758781e-01 3.580380e-05
2.7 -1.705538e-01 3.092788e-05
2.8 -1.654346e-01 2.737854e-05
2.9 -1.605658e-01 2.671698e-05
3 -1.558206e-01 3.270025e-05
3.5 -1.349590e-01 2.474838e-05
4 -1.177997e-01 2.047060e-05
4.5 -1.037276e-01 2.040500e-05
5 -9.209853e-02 2.064973e-05
5.5 -8.231669e-02 2.452686e-05
6 -7.415734e-02 1.774528e-05
6.5 -6.719390e-02 1.722323e-05
7 -6.118632e-02 1.174829e-05
7.5 -5.605387e-02 1.296344e-05
8 -5.154351e-02 1.370652e-05
9 -4.415557e-02 1.197663e-05
10 -3.833344e-02 7.495543e-06
11 -3.368323e-02 8.654728e-06
12 -2.985938e-02 7.984925e-06
13 -2.672607e-02 6.275072e-06
14 -2.407789e-02 6.313480e-06
15 -2.185101e-02 5.878824e-06
16 -1.994120e-02 5.934196e-06
18 -1.683840e-02 4.003073e-06
20 -1.445338e-02 3.297009e-06
22 -1.257125e-02 3.570933e-06
24 -1.106371e-02 2.406277e-06
26 -9.828171e-03 1.909104e-06
28 -8.799607e-03 2.232859e-06
30 -7.936891e-03 1.927775e-06
32 -7.202707e-03 1.683416e-06
36 -6.029717e-03 1.655090e-06
40 -5.135009e-03 1.242002e-06
44 -4.437241e-03 1.089325e-06
48 -3.880961e-03 1.115461e-06
52 -3.428819e-03 8.508322e-07
56 -3.057012e-03 9.008863e-07
60 -2.745095e-03 7.095149e-07
64 -2.482256e-03 6.293888e-07
72 -2.062673e-03 5.098257e-07
80 -1.745511e-03 4.598340e-07
88 -1.501464e-03 3.389754e-07
96 -1.306813e-03 3.204232e-07
104 -1.150206e-03 2.928537e-07
112 -1.021227e-03 2.410436e-07
120 -9.137152e-04 2.243859e-07
128 -8.236340e-04 1.794494e-07
144 -6.804669e-04 1.594421e-07
160 -5.732991e-04 1.260518e-07
176 -4.907903e-04 1.096747e-07
192 -4.258577e-04 9.861806e-08
208 -3.730761e-04 8.137725e-08
224 -3.302606e-04 7.584203e-08
240 -2.946297e-04 6.483560e-08
256 -2.649533e-04 5.915606e-08
288 -2.178939e-04 4.739605e-08
320 -1.827777e-04 3.951203e-08
352 -1.558822e-04 3.291045e-08
384 -1.347720e-04 2.788692e-08
416 -1.177741e-04 2.424459e-08
448 -1.040283e-04 2.099847e-08
480 -9.260722e-05 1.861413e-08
512 -8.305004e-05 1.643573e-08
1000 -2.654239e-05 4.738913e-09
2000 -8.002707e-06 1.290666e-09
5000 -1.603637e-06 2.290388e-10
10000 -4.684482e-07 6.140526e-11
100000 -7.314084e-09 7.501578e-13
\end{lstlisting}

\end{widetext}

\end{document}